\documentclass[prb,aps,amsmath,amssymb,floatfix,reprint]{revtex4-1}
\usepackage{graphicx}
\usepackage{bm}
\usepackage{amsfonts,mathrsfs}
\usepackage{hyperref}
\hypersetup{colorlinks,citecolor=blue,linkcolor=blue}

\begin{document}

\title{The Individual and Collective Effects of Exact Exchange and Dispersion Interactions on the \textit{Ab Initio} Structure of Liquid Water}
\author{Robert A. DiStasio Jr.$^1$}
\author{Biswajit Santra$^1$}
\author{Zhaofeng Li$^2$}
\author{Xifan Wu$^3$}
\author{Roberto Car$^{1,2,4,5}$}
\affiliation{
$^1$Department of Chemistry, Princeton University, Princeton, NJ 08544, USA \\
$^2$Department of Physics, Princeton University, Princeton, NJ 08544, USA \\
$^3$Department of Physics, Temple University, Philadelphia, PA 19122, USA \\
$^4$Princeton Institute for the Science and Technology of Materials, Princeton University, Princeton, NJ 08544, USA \\
$^5$Program in Applied and Computational Mathematics, Princeton University, Princeton, NJ 08544, USA
}

\begin{abstract}

In this work, we report the results of a series of density functional theory (DFT) based \textit{ab initio} molecular dynamics (AIMD) simulations of ambient liquid water using a hierarchy of exchange-correlation (XC) functionals to investigate the individual and collective effects of exact exchange (Exx), \textit{via} the PBE0 hybrid functional, non-local vdW/dispersion interactions, \textit{via} a fully self-consistent density-dependent dispersion correction, and approximate nuclear quantum effects (aNQE), \textit{via} a 30 K increase in the simulation temperature, on the microscopic structure of liquid water.
Based on these AIMD simulations, we found that the collective inclusion of Exx, vdW, and aNQE as resulting from a large-scale AIMD simulation of (H$_2$O)$_{128}$ at the PBE0+vdW level of theory, significantly softens the structure of ambient liquid water and yields an oxygen-oxygen structure factor, $S_{\rm OO}(Q)$, and corresponding oxygen-oxygen radial distribution function, $g_{\rm OO}(r)$, that are now in quantitative agreement with the best available experimental data.
This level of agreement between simulation and experiment as demonstrated herein originates from an increase in the relative population of water molecules in the interstitial region between the first and second coordination shells, a collective reorganization in the liquid phase which is facilitated by a weakening of the hydrogen bond strength by the use of the PBE0 hybrid XC functional, coupled with a relative stabilization of the resultant disordered liquid water configurations by the inclusion of non-local vdW/dispersion interactions.
This increasingly more accurate description of the underlying hydrogen bond network in liquid water obtained at the PBE0+vdW+aNQE level of theory yields higher-order correlation functions, such as the oxygen-oxygen-oxygen triplet angular distribution, $P_{\rm OOO}(\theta)$, and therefore the degree of local tetrahedrality, as well as electrostatic properties, such as the effective molecular dipole moment, that are also in much better agreement with experiment.
%

\end{abstract}

\maketitle

\section{Introduction \label{sec:Intro}}

Water is arguably the most important molecule on Earth. 
Without water, proteins would not fold, salts would not dissolve, and there would certainly not be life, at least as we know it. 
While a single water molecule has a simple and well-known structure, consisting of an oxygen atom covalently bound to two hydrogen atoms, the unique physical and chemical properties of water in the solid and liquid phases are largely due to the subtle interplay between the underlying hydrogen bond network and entropic effects, which collectively govern the interactions between water molecules.
As such, an accurate understanding of the intricate details of the microscopic structure of liquid water---the net result of this competition between hydrogen bond energetics (which favor ordered structures) and temperature/entropic effects (which favor disordered structures)---requires an accurate and balanced theoretical treatment of both the nuclear and electronic degrees of freedom, and has therefore posed a substantial challenge for modern state-of-the-art \textit{ab initio} quantum mechanical methodologies.

A highly accurate and detailed understanding of the microscopic structure of liquid water is of great importance to a number of fields, ranging from biology/biochemistry (\textit{i.e.}, selective ion channel functioning, solvent-mediated protein folding, etc.) to environmental science (\textit{i.e.}, ``green'' solvents, cloud formation, etc.) to energy storage/electrochemistry (\textit{i.e.}, aqueous electrolytes, charge-transfer interactions, catalyst design, etc.).~\cite{franks_book_2000}
While there is no experimental methodology currently available to directly obtain the real-space microscopic structure of liquid water,~\cite{soper_ipc_2013,head-gordon_cr_2002} \textit{ab initio} molecular dynamics (AIMD)~\cite{marx_book_2009} simulations employing Density Functional Theory (DFT)~\cite{burke_jcp_2012} as the source of the underlying quantum mechanical potential can furnish such structural information.
With the AIMD technique, the nuclear potential energy surface is generated ``on the fly'' from the electronic ground state~\cite{car_prl_1985} without the need for empirical input, thereby allowing for a quantum mechanical treatment of not only the structure and dynamics of a given molecular system of interest, but also its electronic and dielectric properties, as well as potential chemical reactions (\textit{i.e.}, the breaking and forming of chemical bonds).
Since the initial pioneering simulations of liquid water,~\cite{laasonen_jcp_1993,tuckerman_jcp_1995} AIMD has been applied to many complex problems in biology, chemistry, and energy research, \textit{e.g.}, the designing of efficient catalysts for hydrogen production~\cite{zipoli_jacs_2010} and the modeling of the autoionization~\cite{geissler_science_2001} and electrocatalytic splitting of water,~\cite{ikeshoji_pccp_2009} to name a few.

In general, the predictive power of DFT-based AIMD simulations depends crucially upon the accuracy of the underlying exchange-correlation (XC) functional utilized in the quantum mechanical treatment of the electronic degrees of freedom.
In current AIMD simulations of liquid water, the most widely used XC functionals involve the generalized gradient approximation (GGA), and it is now evident that the use of GGA-DFT has severe limitations when applied to liquid water~\cite{asthagiri_pre_2003,grossman_jcp_2004,schwegler_jcp_2004,fernandez_jcp_2004,kuo_jpcb_2004,mcgrath_cpc_2005,vandevondele_jcp_2005,sit_jcp_2005,fernandez_ms_2005,mcgrath_mp_2006,lee_jcp_2006,todorova_jpcb_2006,lee_jcp_2007,guidon_jcp_2008,kuhne_jctc_2009,mattson_jctc_2009,yoo_jcp_2009,zhang_jctc_2011-2,bartok_prb_2013,alfe_jcp_2013,lin_jpcb_2009,jonchiere_jcp_2011,wang_jcp_2011,zhang_jctc_2011,mogelhoj_jpcb_2011,lin_jctc_2012,yoo_jcp_2012,schmidt_jpcb_2009,ma_jcp_2012,ben_jpcl_2013} as well as crystalline phases of ice.~\cite{feibelman_pccp_2008,santra_prl_2011,santra_jcp_2013,labat_jcc_2011,kambara_pccp_2012,murray_prl_2012,fang_prb_2013,macher_jcp_2014}
For one, GGA-DFT suffers from the well-known self-interaction error~\cite{perdew_prb_1981} which leads to excessive proton delocalization in liquid water, yielding red shifts in the predicted O-H stretching frequencies as compared to experiment.~\cite{zhang_jctc_2011,zhang_jctc_2011-2}
Secondly, GGA-DFT ignores non-local electron correlation effects that are responsible for van der Waals (vdW) or dispersion interactions, which are thought to be crucial, for instance, in correctly obtaining the equilibrium density of liquid water.~\cite{schmidt_jpcb_2009,wang_jcp_2011,ma_jcp_2012}
Beyond the choice of the XC functional, most AIMD simulations of liquid water performed to date have adopted classical mechanics for the nuclear equations of motion and therefore completely neglect nuclear quantum effects (NQE), and this approximation has also been deemed insufficient for a highly accurate quantitative description of the microscopic structure of liquid water.
In the case of liquid water, light atoms, such as hydrogen in particular, deviate significantly from classical behavior even at room temperature,~\cite{chen_prl_2003,morrone_prl_2008,ceriotti_pnas_2013} and this leads to overstructuring in the predicted radial distribution functions.~\cite{morrone_prl_2008,ceriotti_prl_2012,paesani_jcp_2007,fanourgakis_jcp_2006}

A commonly adopted method to alleviate the self-interaction error in GGA-DFT is the use of hybrid XC functionals, wherein a fraction of the exact exchange (Exx) energy is included in the density functional approximation.
Due to the relatively high computational cost associated with these XC functionals, applications of hybrid-based DFT over the past several years have been mostly restricted to small gas-phase clusters of water,~\cite{santra_jcp_2007,santra_jcp_2008,santra_jcp_2009,wang_jcp_2010,gillan_jcp_2012} although recently hybrid functionals have been applied in the study of several crystalline phases of ice.~\cite{santra_prl_2011,santra_jcp_2013,labat_jcc_2011,kambara_pccp_2012,erba_jpcb_2009}
In comparison to GGAs, these studies demonstrated that the energetic, structural, and vibrational properties of these systems, as predicted by hybrid DFT calculations, are generally in closer agreement with the available experimental data.~\cite{santra_jcp_2009,wang_jcp_2010,gillan_jcp_2012,santra_prl_2011,labat_jcc_2011,kambara_pccp_2012,erba_jpcb_2009,santra_jcp_2013}
Although applications of hybrid functionals to liquid water have been relatively rare in the literature, it was found that PBE0~\cite{perdew_jcp_1996,adamo_jcp_1999} greatly improves upon the accuracy of the vibrational spectrum of liquid water.~\cite{zhang_jctc_2011,zhang_jctc_2011-2}
In the same breath, however, ambiguities exist and still remain concerning the effects of Exx on the structure of liquid water; while some studies have inferred a softening of the structure with hybrid functionals over GGAs,~\cite{zhang_jctc_2011,zhang_jctc_2011-2,todorova_jpcb_2006} others have found the effects of Exx to be negligible in this regard.~\cite{guidon_jcp_2008,ben_jpcl_2013}
As a further complication, the increased computational cost associated with hybrid XC functionals has restricted most of the AIMD simulations performed at this level of theory to small system sizes and/or relatively short simulation times, both of which have the potential to prohibit an accurate assessment of the effects of Exx on the structural properties of liquid water.
With that being said, it is generally accepted that more in-depth studies will be required in order to definitively ascertain the effects of Exx on the microscopic structure of liquid water---and this represents one of the main goals of the research reported in this manuscript.

In addition, both hybrid and GGA XC functionals lack the ability to describe vdW/dispersion interactions, which arise from non-local dynamical electron correlation and have a substantial effect on the microscopic structure of water in both the solid and liquid phases. 
In this regard, the explicit inclusion of pairwise-additive vdW interactions in DFT has been shown to significantly improve upon the theoretical description of the transition pressures among the high-pressure phases of ice~\cite{santra_prl_2011,kambara_pccp_2012,murray_prl_2012} and the predicted equilibrium density of liquid water.~\cite{schmidt_jpcb_2009,ma_jcp_2012,ben_jpcl_2013}
Many recent studies have concluded that the structure of liquid water significantly softens when vdW interactions are accounted for in the underlying XC potential; however, the extent to which these non-local vdW forces affect the structure of liquid water is largely dependent upon the given approach utilized to facilitate vdW-inclusive DFT.~\cite{lin_jpcb_2009,jonchiere_jcp_2011,wang_jcp_2011,zhang_jctc_2011,mogelhoj_jpcb_2011,lin_jctc_2012,yoo_jcp_2012,schmidt_jpcb_2009,ma_jcp_2012,ben_jpcl_2013}
For instance, the pairwise-additive $C_6/R^6$ based vdW correction approaches of Grimme,~\cite{grimme_jcc_2004,grimme_jcp_2010} when used in conjunction with the popular PBE~\cite{perdew_prl_1996} and BLYP~\cite{becke_pra_1988,lee_prb_1988} GGA functionals, tend to reduce the intensity of the first maximum in the oxygen-oxygen radial distribution function ($g_{\rm OO}(r)$) of liquid water over a fairly wide range, \textit{i.e.}, by approximately 5\% for PBE-D~\cite{schmidt_jpcb_2009} and 10-17\% for BLYP-D.~\cite{lin_jpcb_2009,schmidt_jpcb_2009,jonchiere_jcp_2011}
Other popular vdW-inclusive DFT approaches include the so-called vdW-DF functionals,~\cite{dion_prl_2004} which incorporate vdW interactions \textit{via} the use of explicit non-local correlation functionals, have demonstrated severe shortcomings in reproducing the second coordination shell in ambient liquid water.~\cite{wang_jcp_2011,mogelhoj_jpcb_2011}
Again, an accurate assessment of the effects of non-local vdW/dispersion interactions on the microscopic structure of liquid water, in particular when these interactions are treated in conjunction with a hybrid XC functional, represents an important step in increasing our understanding of this fundamental aqueous system, and this indeed is the main focus of the research reported herein.

In order to understand and quantify the individual and collective effects of Exx and vdW on the microscopic structure of liquid water, we have systematically performed a series of AIMD simulations of liquid water at ambient conditions using a hierarchy of XC functionals with increasing accuracy.
In particular, the sequence of XC functionals employed in this work includes the standard semi-local GGA of Perdew, Burke, and Ernzerhof~\cite{perdew_prl_1996} (PBE), the corresponding PBE0~\cite{perdew_jcp_1996,adamo_jcp_1999} hybrid, and self-consistent dispersion-corrected analogs~\cite{distasio_unpublished} of each, \textit{i.e.}, PBE+vdW and PBE0+vdW, based on the Tkatchenko-Scheffler~\cite{tkatchenko_prl_2009} density-dependent vdW/dispersion functional (TS-vdW). 
To successfully perform these AIMD simulations, we have employed a linear scaling O($N$) algorithm to compute the exact exchange energy,~\cite{wu_prb_2009} which allows for relatively long ($t_{sim} > 25$ ps) hybrid functional based simulations using large system sizes (\textit{e.g.}, the explicit treatment of 128 water molecules).
In addition, we have also developed and utilized a linear scaling O($N$) self-consistent implementation of the TS-vdW dispersion correction, which provides a framework for computing atomic $C_6$ dispersion coefficients as an explicit functional of the charge density, thereby accounting for the local chemical environment surrounding each atom.~\cite{tkatchenko_prl_2009}
In this manuscript, we demonstrate that the individual and collective effects of Exx and vdW interactions significantly soften the structure of liquid water and predict structural properties that are in closer agreement with the available experimental data as compared to the now well-established predictions of the PBE GGA functional.
Since it is also now known that a classical treatment of the nuclear degrees of freedom has been deemed insufficient for obtaining a quantitatively accurate description of the microscopic structure of liquid water, we have also performed an additional AIMD simulation at the slightly elevated temperature of 330 K, a technique which has been shown to mimic the NQE in the $g_{\rm OO}(r)$.~\cite{morrone_prl_2008,paesani_jcp_2007}
By approximately accounting for NQE in AIMD simulations at the PBE0+vdW level of theory, which is the most accurate XC functional considered in this work, we have obtained quantitative agreement with scattering experiments in the $g_{\rm OO}(r)$ and significant systematic improvements in many other structural properties (\textit{i.e.}, the oxygen-hydrogen radial distribution function, local tetrahedrality and the oxygen-oxygen-oxygen angular distribution function, characterization of the underlying hydrogen bonding network) as well as electrostatic properties (such as the molecular dipole moment) of liquid water.

The remainder of the manuscript is organized as follows. 
In Sec.~\ref{sec:SimulationDetails}, we describe the computational details of the AIMD simulations performed herein.
In Sec.~\ref{sec:Theory}, we present the theoretical methodologies developed and utilized in this work to study the structural properties of liquid water at ambient conditions.
Sec.~\ref{sec:ResultsandDiscussion} contains an in-depth discussion of the results of these simulations and a comparative analysis with the currently available theoretical and experimental literature on ambient liquid water.
The paper is then completed with Sec.~\ref{sec:Conclusions}, which provides some brief conclusions as well as the future outlook of AIMD simulations of liquid water.

\section{Simulation Details\label{sec:SimulationDetails}}

In this work, we have systematically performed DFT-based AIMD simulations of ambient liquid water using a hierarchy of different XC functionals.
The sequence of XC functionals employed herein includes the standard semi-local PBE-GGA,~\cite{perdew_prl_1996} the corresponding hybrid PBE0~\cite{perdew_jcp_1996,adamo_jcp_1999} which includes 25\% exact exchange, and the self-consistent dispersion-corrected analogs~\cite{distasio_unpublished} thereof, \textit{i.e.}, PBE+vdW and PBE0+vdW, based on the Tkatchenko-Scheffler~\cite{tkatchenko_prl_2009} density-dependent vdW/dispersion functional. 
Further details of the theoretical methods employed in this work are provided directly below in Sec.~\ref{sec:Theory}.

Each of these AIMD simulations of liquid water was performed in the canonical ($NVT$) ensemble using periodic simple cubic simulation cells with lattice parameters set to reproduce the experimental density of liquid water at ambient conditions.
The Car-Parrinello (CP)~\cite{car_prl_1985} equations of motion for the nuclear and electronic degrees of freedom were integrated using the standard Verlet algorithm and a time step of 4.0 a.u. ($\approx$ 0.1 fs).
The ionic temperatures were controlled with Nos\'{e}-Hoover chain thermostats, each with a chain length of 4.~\cite{martyna_jcp_1992}
To achieve rapid equipartition of the thermal energy, we employed one Nos\'{e}-Hoover chain thermostat per atom (\textit{i.e.}, the so-called ``massive'' Nos\'{e}-Hoover thermostat) and also rescaled the fictitious thermostat masses by the atomic masses, so that the relative rates of the thermostat fluctuations were inversely proportional to the masses of the atoms to which they were coupled.~\cite{tobias_jpc_1993}
The electronic wavefunctions were expanded using a plane wave basis set with a kinetic energy cutoff of 71 Ry.
The interactions between the valence electrons and the ions (consisting of the nuclei and their corresponding frozen-core electrons) were treated with Troullier-Martins type norm-conserving pseudopotentials.~\cite{troullier_prb_1991}
To ensure an adiabatic separation between the electronic and nuclear degrees of freedom in the CP dynamics, we used a fictitious electronic mass of 300 a.u., which was found to be a reasonable choice for the simulation of water,~\cite{grossman_jcp_2004} and the nuclear mass of deuterium for each hydrogen atom.
Mass preconditioning was applied to all Fourier components of the electronic wavefunctions having a kinetic energy greater than 3 Ry.~\cite{tassone_prb_1994}

Using a simulation cell with $L=12.4$ \AA, four AIMD simulations were performed on (H$_2$O)$_{64}$ at 300 K using the PBE, PBE0, PBE+vdW, and PBE0+vdW XC functionals. 
Since it is now well-known that a classical treatment of the nuclear degrees of freedom is insufficient for a quantitatively accurate description of the microscopic structure of liquid water at room temperature, we have also performed an additional AIMD simulation on (H$_2$O)$_{128}$ with a temperature of 330 K at the PBE0+vdW level, which is the most accurate XC functional considered herein. 
In this regard, an increase of approximately 30 K in the simulation temperature has been shown to mimic the nuclear quantum effects (NQE) in structural quantities such as the oxygen-oxygen radial distribution function ($g_{\rm OO}(r)$) in both DFT~\cite{morrone_prl_2008} and force-field~\cite{paesani_jcp_2007,fanourgakis_jcp_2006} based MD simulations.

All calculations reported in this work utilized a modified development version of the Quantum ESPRESSO (QE) software package.~\cite{QE-2009}  
All of the simulations were initially equilibrated for approximately 2 ps and then continued for at least an additional 20 ps for data collection.
The largest AIMD simulation performed herein (\textit{i.e.}, (H$_2$O)$_{128}$ at the PBE0+vdW level of theory) employed 1024 cores on the Cray XE6 platform at the National Energy Research Scientific Computing Center (NERSC) for $\approx$ 5 days to carry out 1 ps of simulation.

\section{Theoretical Methods \label{sec:Theory}}

\subsection{Hybrid XC Functionals \label{sec:PBE0}}

The utilization of hybrid XC functionals requires significant additional computational cost relative to GGA-DFT-based AIMD simulations.
To meet these additional computational demands and thereby allow for large-scale AIMD simulations based on hybrid XC functionals, we have employed a linear scaling O($N$) exact exchange algorithm, which has been developed~\cite{wu_prb_2009} and extensively optimized in our research group. 
The computational savings afforded by this algorithm originate from the maximally localized Wannier function (MLWF)~\cite{marzari_prb_1997} representation of the occupied subspace, $\{\widetilde{\varphi}\}$, obtained \textit{via} a unitary transformation of the occupied Kohn-Sham electronic states, $\{\varphi\}$, \textit{i.e.}, $\widetilde{\varphi}_i=\sum_j U_{ij}\varphi_j$, and the subsequent efficient use of sparsity in the numerical evaluation of all required quantities in real-space.

In this approach, the (closed-shell) exact exchange energy,
\begin{equation}
\label{eq:exx}
E_{\rm xx}=-2\sum_{ij}\int d\mathbf{r}\,\widetilde{\varphi}_i^{*}(\mathbf{r})\widetilde{\varphi}_j(\mathbf{r})v_{ij}(\mathbf{r}) ,
\end{equation}
and orbital-dependent ``force'' acting on the electronic wavefunctions in the CP dynamics,~\cite{car_prl_1985}
\begin{equation}
\label{eq:hxx}
D_{\rm xx}^i(\mathbf{r})=-\left(\frac{\delta E_{\rm xx}}{\delta \widetilde{\varphi}_i^{*}(\mathbf{r})}\right)=2\sum_{j}\widetilde{\varphi}_j(\mathbf{r})v_{ij}(\mathbf{r}) ,
\end{equation}
can both be written in terms of $v_{ij}(\mathbf{r})$, the electrostatic potential generated by the pair density, $\widetilde{\rho}_{ij}(\mathbf{r})=\widetilde{\varphi}_i(\mathbf{r})\widetilde{\varphi}_j^{*}(\mathbf{r})$, \textit{i.e.}, 
\begin{equation}
\label{eq:vxx}
v_{ij}(\mathbf{r})=\int d\mathbf{r'}\,\frac{\widetilde{\rho}_{ij}(\mathbf{r'})}{|\mathbf{r}-\mathbf{r'}|} .
\end{equation}
The potential $v_{ij}(\mathbf{r})$ is the solution of the Poisson equation, 
\begin{equation}
\label{eq:pe}
\nabla^2v_{ij}(\mathbf{r})=-4\pi\widetilde{\rho}_{ij}(\mathbf{r}) ,
\end{equation}
subject to the boundary condition that $v_{ij}(\mathbf{r}) \rightarrow 0$ at infinity.
By virtue of the localized character of $\widetilde{\rho}_{ij}(\mathbf{r})$, the potential $v_{ij}(\mathbf{r})$ at sufficient distances from the center of the pair density is exactly represented by a multipolar expansion,
\begin{equation}
\label{eq:me}
v_{ij}(\mathbf{r})=4\pi\sum_{lm}\frac{Q_{lm}}{(2l+1)}\frac{Y_{lm}(\theta,\phi)}{r^{l+1}} ,
\end{equation}
in which $Q_{lm}=\int d\mathbf{r}\,Y_{lm}^{*}(\theta,\phi)r^l\widetilde{\rho}_{ij}(\mathbf{r})$ are the multipoles of the charge distribution $\widetilde{\rho}_{ij}(\mathbf{r})$.
The integrals required to evaluate the multipoles are restricted to a volume that fully contains $\widetilde{\rho}_{ij}(\mathbf{r})$, as delimited by the Poisson equation cutoff radius, \textit{i.e.}, for $|\mathbf{r}| < R_{PE}$.
Eq.~(\ref{eq:pe}) is then solved numerically in real-space (\textit{i.e.}, on the dense real-space grid) within the sphere defined by $R_{PE}$, with the boundary condition provided by Eq.~(\ref{eq:me}) for $|\mathbf{r}| = R_{PE}$.

To efficiently compute the exact exchange energy at every AIMD step, the required MLWFs were evaluated by minimizing the total spread functional, $S[\widetilde{\varphi}_i(\mathbf{r})]$, using second-order damped CP dynamics as described in Ref.~[\onlinecite{sharma_ijqc_2003}].
During this procedure, we used a fictitious mass of 500 a.u., a damping coefficient of 0.3, and a time step of 4 a.u.
On average, less than 5 iterations per AIMD step were sufficient to reach a convergence of 10$^{-6}$ in the optimal spread $S[\widetilde{\varphi}_i(\mathbf{r})]$.
The resulting average spread for a single MLWF was $\approx$ 0.52 \AA$^2$ during the PBE0 and PBE0+vdW simulations performed in this work.
%
For a given pair of occupied Kohn-Sham electronic states, the exact exchange energy was computed when the corresponding MLWF centers were within 3.70 \AA. 
The Poisson equation ($R_{PE}$) cutoff radius above was defined with respect to the center of a given pair density $\widetilde{\rho}_{ij}(\mathbf{r})$ and set to a value of 2.65 \AA, to ensure an accurate computation of the exact exchange energy. 
The maximum angular momentum component retained in the multipolar expansion given in Eq.~(\ref{eq:me}) was $l_{max}=6$, which has previously been found to be sufficiently accurate in computing exact exchange energies in a variety of systems.~\cite{wu_prb_2009}
This procedure provides the exact exchange energy with an accuracy of at least $10^{-4}$ Hartrees/molecule, as was tested by systematically increasing the $R_{PE}$ cutoff radius and including pairs of MLWFs up to distances of half the simulation cell length.

\subsection{Self-Consistent Dispersion-Corrected XC Functionals \label{sec:SCvdW}}

To account for the non-local vdW/dispersion interactions in our AIMD simulations of liquid water, we developed a fully self-consistent O($N$) implementation~\cite{distasio_unpublished} of the density-dependent dispersion correction of Tkatchenko and Scheffler~\cite{tkatchenko_prl_2009} (TS-vdW) into a development version of QE.
The TS-vdW energy is constructed as an explicit functional of the charge density and is written as a sum over atomic pair energies for atoms $A$ and $B$ as
\begin{equation}
E_{\rm vdW}[\rho(\mathbf{r})]=-\frac{1}{2}\sum_{AB}\frac{f_{AB}[\rho(\mathbf{r})]C_{6,AB}[\rho(\mathbf{r})]}{R_{AB}^{6}} ,
\end{equation}
in which $f_{AB}[\rho(\mathbf{r})]$ is a Fermi-type damping function (used to ensure that the vdW correction goes continuously to zero at short interatomic distances) that explicitly depends on the charge density through the vdW radii of atoms $A$ and $B$, $f_{AB}[\rho(\mathbf{r})]=f_{AB}(R_{AB},R_A^0[\rho(\mathbf{r})],R_B^0[\rho(\mathbf{r})])$, $C_{6,AB}[\rho(\mathbf{r})]$ is the effective heteronuclear dispersion coefficient that is also dependent on the charge density \textit{via} the static dipole polarizabilities and homonuclear dispersion coefficients of atoms $A$ and $B$, $C_{6,AB}[\rho(\mathbf{r})]=C_{6,AB}(\alpha_A^0[\rho(\mathbf{r})],\alpha_B^0[\rho(\mathbf{r})],C_{6,AA}[\rho(\mathbf{r})],C_{6,BB}[\rho(\mathbf{r})])$, and $R_{AB}$ is the scalar distance between atoms $A$ and $B$, $R_{AB}=|\mathbf{R}_A-\mathbf{R}_B|$.
As such, the TS-vdW correction provides a theoretical framework which accounts for local chemical environment effects, such as hybridization, exchange-correlation, and Pauli repulsion, in the calculation of the vdW energy and forces, and is therefore not strictly pairwise-additive~\cite{dobson_ijqc_2014} as \textit{e.g.}, the DFT+D approaches of Grimme.~\cite{grimme_jcc_2004,grimme_jcp_2010}

The implementation utilized herein in conjunction with the PBE and PBE0 XC functionals, namely PBE+vdW and PBE0+vdW, respectively, allows us to self-consistently account for the changes in the total energy and charge density ($E[\rho(\mathbf{r})]=E_{\rm DFT}[\rho(\mathbf{r})]+E_{\rm vdW}[\rho(\mathbf{r})]$) that arise from the fact that the vdW/dispersion correction itself is an explicit correlation functional of the charge density.
The fully self-consistent implementation of this energy expression for use in AIMD simulations therefore requires a corresponding electronic potential and ``force'' acting on the electronic wavefunctions in the CP dynamics,~\cite{car_prl_1985}
\begin{equation}
\label{eq:hvdw}
D_{\rm vdW}^i(\mathbf{r})=-\left(\frac{\delta E_{\rm vdW}}{\delta \varphi_i^{*}(\mathbf{r})}\right)=-V_{\rm vdW}(\mathbf{r})\varphi_i(\mathbf{r}) ,
\end{equation}
in which $V_{\rm vdW}(\mathbf{r})$ is the local vdW/dispersion potential derived \textit{via} $V_{\rm vdW}(\mathbf{r})=\left(\frac{\delta E_{\rm vdW}}{\delta \rho(\mathbf{r})}\right)$ as
\begin{eqnarray}
V_{\rm vdW}(\mathbf{r})=-\frac{1}{2}\sum_{AB}&\Biggl\{&\frac{\left(\frac{\delta f_{AB}[\rho(\mathbf{r})]}{\delta\rho(\mathbf{r})}\right) C_{6,AB}[\rho(\mathbf{r})]}{R_{AB}^{6}} \nonumber \\
&+&\frac{f_{AB}[\rho(\mathbf{r})] \left(\frac{\delta C_{6,AB}[\rho(\mathbf{r})]}{\delta\rho(\mathbf{r})}\right)}{R_{AB}^{6}}\Biggr\} ,
\end{eqnarray}
which is added to the local XC potential, $V_{\rm XC}(\mathbf{r})$, at each MD step.

During the AIMD simulations at the PBE+vdW and PBE0+vdW levels of theory, the $s_R$ scaling parameter in the damping function above, which determines the onset of the vdW correction for a given XC functional, was set to 0.94 and 0.96, respectively, as recommended in Ref.~[\onlinecite{tkatchenko_prl_2009}].
The free atom reference quantities utilized herein were $\alpha_{\rm O}^0=5.4$ bohr$^3$, $C_{6,\rm OO}=15.6$ Hartree $\cdot$ bohr$^6$, and $R_{\rm O}^0=3.19$ bohr for oxygen and $\alpha_{\rm O}^0=4.5$ bohr$^3$, $C_{6,\rm OO}=6.5$ Hartree $\cdot$ bohr$^6$, and $R_{\rm O}^0=3.10$ bohr for hydrogen.
The vdW energy (and resultant ionic forces and electronic potential) were computed in real-space by explicitly summing over the atoms in the simulation and neighboring periodic cells to ensure convergence in $E_{\rm vdW}$ to at least $10^{-5}$ Hartrees/molecule.

\section{Results and Discussion \label{sec:ResultsandDiscussion}}

\subsection{Comparative Analysis of the Oxygen-Oxygen Structure Factor \label{sec:OOSF}}

\begin{figure}
\begin{center}
\includegraphics[width=8.75cm]{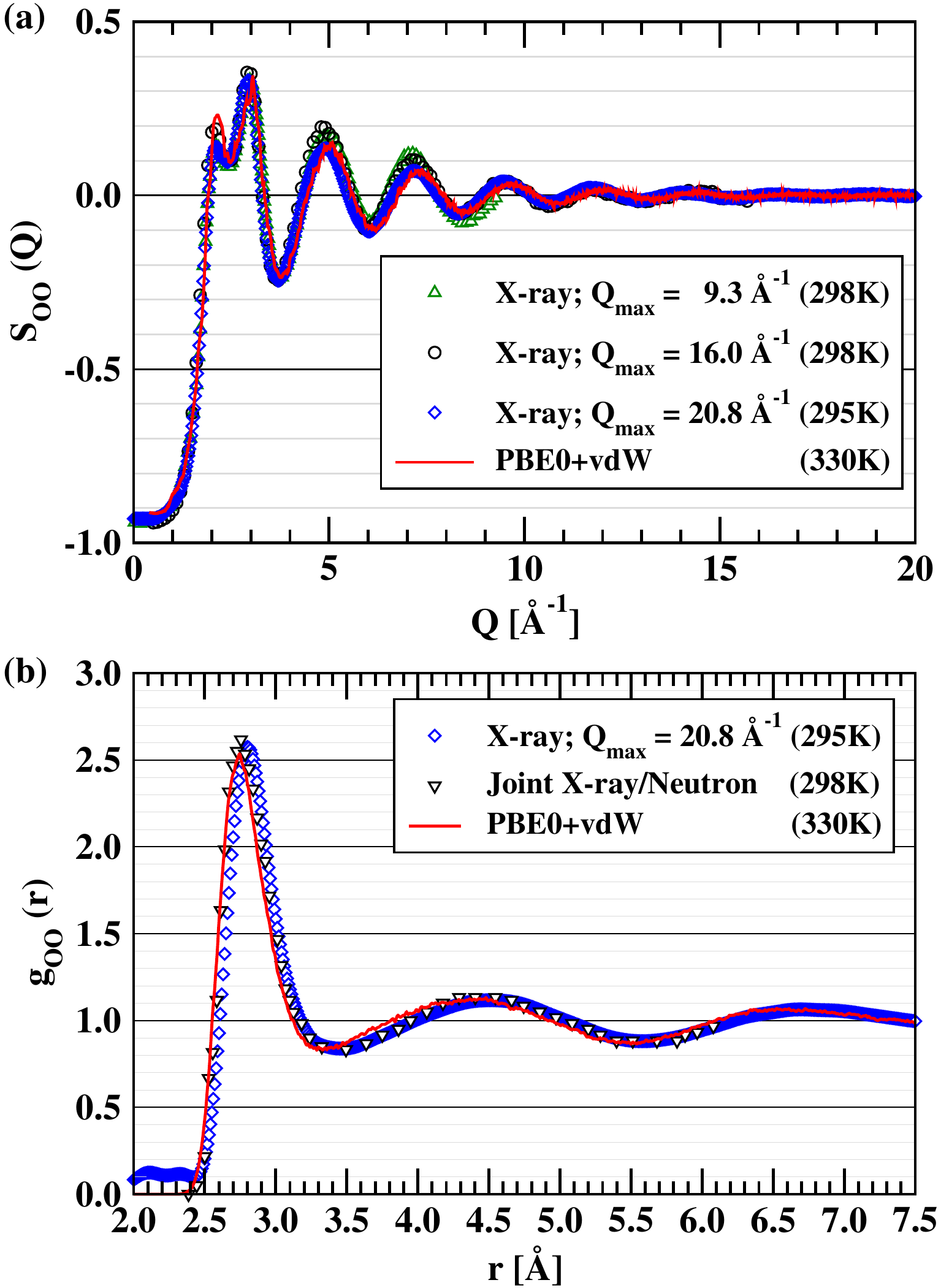}
\caption{\label{fig:gOOSOO} \textbf{Panel (a)}: The oxygen-oxygen structure factors, $S_{\rm OO}(Q)$, of liquid water obtained from theory (\textit{via} the DFT-based AIMD simulations performed in this work) and various X-ray scattering experiments.~\cite{sorenson_jcp_2000,huang_pccp_2011,skinner_jcp_2013} \textbf{Panel (b)}: The oxygen-oxygen radial distribution functions, $g_{\rm OO}(r)$, of liquid water obtained from theory (\textit{via} the DFT-based AIMD simulations performed in this work) and various scattering experiments.~\cite{skinner_jcp_2013,soper_prl_2008}}
\end{center}
\end{figure}

We begin our analysis by comparing the oxygen-oxygen structure factor, $S_{\rm OO}(Q)$, obtained from the highest quality AIMD simulations performed herein at the PBE0+vdW (330 K) level of theory and various X-ray scattering measurements (See Fig.~\ref{fig:gOOSOO}(a)).
Such X-ray scattering experiments directly probe the electron charge density (which is centered on the oxygen atoms in the case of water), thereby allowing for $S_{\rm OO}(Q)$ to be accurately obtained from the directly measured scattering intensity, $I(Q)$, and knowledge of the corresponding form factor, $F^2(Q)$, \textit{via} $S(Q)=I(Q)/F^2(Q)$, in which $Q$ represents the change in the wavevector between the incident and scattered radiation. 
%
Once $S(Q)$ has been experimentally determined, the so-called pair correlation or radial distribution function, $g(r)$, which is the most commonly utilized measure of the microscopic structure of a liquid in real space, can then be obtained \textit{via} the inverse Fourier transform, \textit{i.e.},
\begin{equation}
g(r)=1+\frac{1}{2\pi^2\rho r}\int_0^{Q_{max}} dQ\,Q[S(Q)-1]{\rm sin}(Qr) ,
\end{equation}
wherein $Q_{max}$ is the maximum value of $Q$ in the given scattering experiment and $\rho$ is the atomic number density.
Hence, the process of obtaining an accurate $g(r)$ is highly sensitive to the $Q_{max}$ utilized in a given X-ray scattering experiment, which still represents the main technical challenge for scattering experiments to date, and the \textit{a priori} knowledge of the corresponding form factor, $F^2(Q)$, required for deriving $S(Q)$ from the experimentally measured scattering intensity.

As shown in Fig.~\ref{fig:gOOSOO}(a), the $S_{\rm OO}(Q)$ taken from various X-ray scattering experiments~\cite{sorenson_jcp_2000,huang_pccp_2011,skinner_jcp_2013} with increasing values of $Q_{max}$ are very similar across the range of $Q$ values accessible in a given scattering experiment; despite this fact, each of these X-ray scattering experiments yields significantly different $g_{\rm OO}(r)$, especially with respect to the intensity and position of the first peak.~\cite{sorenson_jcp_2000,huang_pccp_2011,skinner_jcp_2013}
In this regard, we note that these differences in the experimentally derived $g_{\rm OO}(r)$ are almost entirely due to the aforementioned $Q_{max}$ limitation inherent to the X-ray scattering experiments and not due to errors in the corresponding $F^2(Q)$.
To justify this claim, we first computed the difference between the $F^2(Q)$ most commonly utilized in these X-ray scattering experiments, which are obtained from high-level quantum chemistry calculations on a single gas-phase water molecule,~\cite{wang_jcp_1994,watnabe_jcp_1999} and the corresponding quantity at the DFT level of theory.
Finding nearly quantitative agreement in this quantity at both levels of theory for a single gas-phase water molecule, as also demonstrated by Krack \textit{et al.},~\cite{krack_jcp_2002} we then computed the difference between the DFT-based $F^2(Q)$ for the single gas-phase water molecule and the $F^2(Q)$ corresponding to condensed-phase liquid water obtained from the DFT-based AIMD simulations performed herein.~\cite{zhaofeng_thesis_2012}
In performing this analysis, we found that the differences between the form factors computed with respect to gas- and condensed-phase water are extremely small (\textit{i.e.}, to within 1-2\%) and are mainly concentrated in the low-$Q$ region (\textit{i.e.}, for $Q < 2$ \AA$^{-1}$); hence, such differences would be essentially negligible when computing the corresponding $g_{\rm OO}(r)$.

Using an AIMD trajectory obtained at the PBE0+vdW (330 K) level of theory, we have computed $S_{\rm OO}(Q)$ as:
\begin{equation}
S_{\rm OO}(Q)=\frac{1}{NN_Q} \sum_{i,j} \sum_{|\mathbf Q|} {\rm exp}[i\mathbf Q\cdot(\mathbf R_i - \mathbf R_j)] ,
\end{equation}
in which $N$ is the number of oxygen atoms in the simulation cell, $N_Q$ is the degeneracy of the wavevector shell of length $Q=|\mathbf Q|$, and $\mathbf R_i$ is the real-space position vector of $i$-th oxygen atom.
Unlike the experimentally derived $S_{\rm OO}(Q)$, the accuracy of the computed $S_{\rm OO}(Q)$ in the low-$Q$ region is subject to a large degree of numerical noise, a consequence of the fact that there is a relatively small number of $\mathbf Q$ vectors present in this sector.
To combat this fact (and to minimize finite size effects), we performed the highest quality AIMD simulations in this work (at the PBE0+vdW (330 K) level) using a periodic simple cubic box containing 128 water molecules ($L = 15.7$ \AA), wherein the smallest accessible $Q$ value is 0.4 \AA$^{-1}$, which helps in improving the accuracy of the theoretically determined $S_{\rm OO}(Q)$ in the low-$Q$ region.

From Fig.~\ref{fig:gOOSOO}(a), it is clear that the theoretically determined $S_{\rm OO}(Q)$ is in nearly quantitative agreement with the experimental results across the entire $Q$ region accessible to the X-ray scattering experiments, with only a slightly noticeable shift towards higher $Q$ values.
This level of agreement between theory and experiment has been achieved \textit{via} the utilization of the PBE0+vdW XC functional in AIMD simulations performed at 330 K---a level of theory which (\textit{i}) reduces the level of self-interaction with respect to standard GGA-DFT by including 25\% exact exchange (Exx), (\textit{ii}) accounts for non-local vdW/dispersion interactions by a self-consistent density-dependent $C_6/R^6$ correction, and (\textit{iii}) approximately corrects for nuclear quantum effects (aNQE) with a 30 K increase in the simulation temperature (See Sec.~\ref{sec:Theory} for more details on the theoretical methodologies employed in this work).
In what follows, we will focus on the corresponding $g_{\rm OO}(r)$, a real-space quantity which allows for a more straightforward comparative analysis and delineation of the individual and collective effects stemming from each of these improvements in the underlying XC functional.

\subsection{Comparative Analysis of the Oxygen-Oxygen Radial Distribution Function \label{sec:OORDF}}

Within the last year, it has been shown that the accuracy of the oxygen-oxygen radial distribution function, $g_{\rm OO}(r)$, directly obtained from X-ray scattering experiments can be systematically improved upon \textit{via} the use of larger $Q_{max}$ values, which has yielded essentially converged results (\textit{i.e.}, to within 1\%) for the intensity and position of the first peak in the $g_{\rm OO}(r)$ of liquid water with $Q_{max} > 20.0$ \AA$^{-1}$.~\cite{skinner_jcp_2013}
In Fig.~\ref{fig:gOOSOO}(b), this latest X-ray scattering data,~\cite{skinner_jcp_2013} which is now in quantitative agreement with the $g_{\rm OO}(r)$ obtained from empirical potential structural refinement (EPSR) based on joint X-ray/neutron data by Soper and Benmore,~\cite{soper_prl_2008} are both compared against the highest quality AIMD simulations performed herein at the PBE0+vdW (330 K) level of theory.
From this figure, one immediately notices that the $g_{\rm OO}(r)$ obtained from PBE0+vdW (330 K) AIMD simulations is in excellent agreement with both of the experimentally derived results, a key finding of this work which will be discussed in greater detail below.

\begin{figure}
\begin{center}
\includegraphics[width=8.75cm]{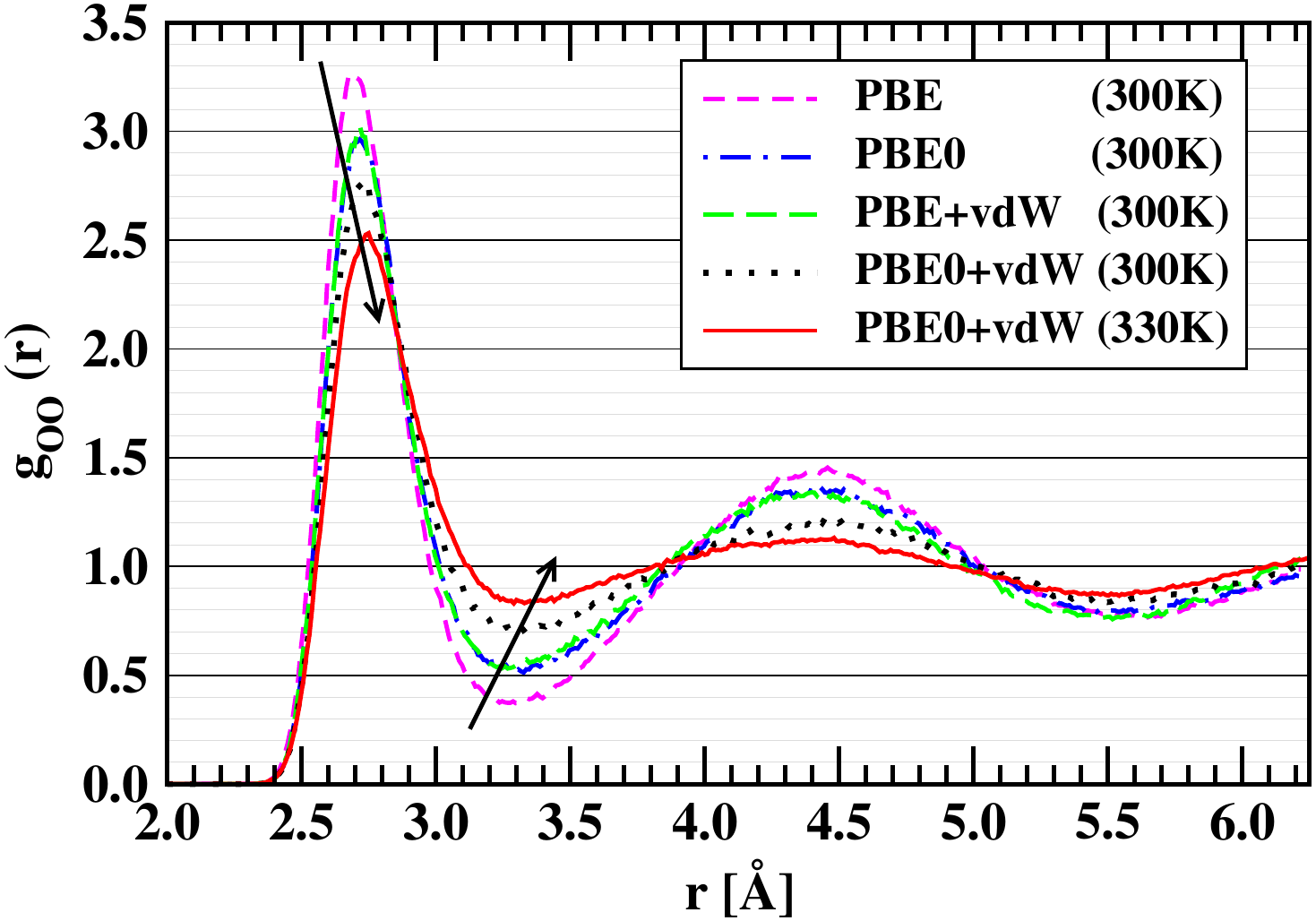}
\caption{\label{fig:gOO} The oxygen-oxygen radial distribution functions $g_{\rm OO}(r)$ of liquid water obtained from theory (\textit{via} the DFT-based AIMD simulations performed in this work). The arrows indicate systematic shifts in the main peak positions and intensities of the computed distributions with improvement of the underlying exchange-correlation functional.}
\end{center}
\end{figure}

As mentioned above, this level of nearly quantitative agreement with experiment has been achieved \textit{via} the collective treatment of exact exchange (\textit{via} the PBE0 hybrid functional), non-local vdW/dispersion interactions (\textit{via} a self-consistent density-dependent $C_6/R^6$ correction) and approximate nuclear quantum effects (\textit{via} a 30 K increase in the simulation temperature).
To illustrate this point, we will now compare the individual and collective effects of Exx, vdW, and aNQE on the predicted $g_{\rm OO}(r)$ of liquid water by considering Fig.~\ref{fig:gOO}, which provides the $g_{\rm OO}(r)$ results obtained from AIMD simulations based on each of the different XC functionals utilized in this work and Table~\ref{tab1}, which summarizes the positions and intensities at various key points in each $g_{\rm OO}(r)$, the oxygen-oxygen coordination numbers, $n_{\rm OO}$, associated with the first shell in liquid water, as well as other structural properties which will be discussed in detail in Sec.~\ref{sec:HB} - \ref{sec:DM}.
\begin{table*}
\caption{\label{tab1} 
Tabulated summary of the structural properties of liquid water obtained \textit{via} the DFT-based AIMD simulations performed in this work and various scattering experiments. From \textit{left to right}, the specific XC functional and temperature ($T$ in K) utilized in a given AIMD simulation; positions (in \AA) and intensities of the first maximum ($r^{\rm max}_{1}$ and $g^{\rm max}_{1}$), the first minimum ($r^{\rm min}_{1}$ and $g^{\rm min}_{1}$), and the second maximum ($r^{\rm max}_{2}$ and $g^{\rm max}_{2}$) of the corresponding oxygen-oxygen radial distribution functions $g_{\rm OO}(r)$ (See Sec.~\ref{sec:OORDF}); oxygen-oxygen coordination numbers ($n_{\rm OO}$) associated with the first shell, computed by integrating each respective $g_{\rm OO}(r)$ up to the first minimum (See Sec.~\ref{sec:OORDF}); the average tetrahedrality parameter ($q$) per water molecule (See Sec.~\ref{sec:OOO}); the average number of hydrogen bonds ($n_{HB}$) per water molecule (See Sec.~\ref{sec:HB}); the average local dipole moment ($\mu$ in Debye) per water molecule (See Sec.~\ref{sec:DM}). For reference, the available corresponding experimental data is provided in the last two rows.}
\begin{ruledtabular}
\begin{tabular}{cccccccccccc}
  Method & $T$ & $r^{\rm max}_{1}$ & $g^{\rm max}_{1}$ & $r^{\rm min}_{1}$ & $g^{\rm min}_{1}$ & $r^{\rm max}_{2}$ & $g^{\rm max}_{2}$ & $n_{\rm OO}$ & $n_{HB}$ & $q$  & $\mu$ \\ 
\hline        
PBE      & 300  & 2.69              & 3.28              & 3.28              & 0.37              & 4.44              & 1.44              & 4.03     & 3.83     & 0.78 &  3.19 \\
PBE+vdW  & 300  & 2.71              & 2.99              & 3.27              & 0.54              & 4.40              & 1.33              & 4.01     & 3.74     & 0.74 &  3.12 \\
PBE0     & 300  & 2.71              & 2.96              & 3.30              & 0.53              & 4.39              & 1.35              & 4.10     & 3.71     & 0.73 &  3.00 \\
PBE0+vdW & 300  & 2.72              & 2.76              & 3.31              & 0.70              & 4.47              & 1.20              & 4.22     & 3.62     & 0.69 &  2.96 \\
PBE0+vdW & 330  & 2.74              & 2.51              & 3.33              & 0.84              & 4.44              & 1.22              & 4.32     & 3.48     & 0.64 &  2.91 \\
\hline            
Expt. 1~\cite{skinner_jcp_2013} & 295  & 2.80 & 2.57     & 3.45              & 0.84              & 4.50              & 1.12              & 4.3(1)   & -        & -    & -     \\
Expt. 2~\cite{soper_prl_2008}   & 298  & 2.76 & 2.62     & 3.42              & 0.82              & 4.43              & 1.14              & 4.67(5)  & -        & 0.58 & -     \\
\end{tabular}
\end{ruledtabular}
\end{table*}

The first point to note here is the fact that the $g_{\rm OO}(r)$ obtained from the PBE based AIMD simulations performed herein are in good agreement with the previous reports in the literature, wherein the intensity of the first maximum, $g^{\rm max}_{1}$ is typically found to be greater than 3.2,~\cite{grossman_jcp_2004,schwegler_jcp_2004,kuhne_jctc_2009} followed by a first minimum that is too deep with respect to experiment.
This level of overstructuring in the $g_{\rm OO}(r)$ of liquid water generated at the PBE level of theory is fairly well-known at this point and is a manifestation of the key limitations of GGA-DFT in the theoretical description of liquid water, namely the presence of self-interaction error and the lack of non-local vdW/dispersion interactions in the underlying XC potential.
In other words, these inherent limitations in the underlying XC potential essentially yield a system that is more representative of deeply supercooled liquid water when the corresponding PBE-based AIMD simulations are performed at ambient conditions.

By including a fraction (\textit{i.e.}, 25\%) of exact exchange \textit{via} the use of the PBE0 hybrid functional, an immediate softening of the $g_{\rm OO}(r)$ is observed in which the intensity of the first maximum, $g^{\rm max}_{1}$, is reduced by approximately 0.32 and the intensity of the first minimum, $g^{\rm min}_{1}$, is increased by approximately 0.16, coupled with an increase in the positions of the first maximum, $r^{\rm max}_{1}$, and first minimum, $r^{\rm min}_{1}$, by 0.02 \AA\, with respect to the $g_{\rm OO}(r)$ generated at the PBE level of theory (See Fig.~\ref{fig:gOO} and Table~\ref{tab1}).
These noticeable differences in the $g_{\rm OO}(r)$ are primarily due to the fact that the PBE0 hybrid functional, which reduces the amount of self-interaction error in the underlying XC potential and thereby corrects for the overly delocalized electron density at the GGA-DFT level of theory, weakens the hydrogen bond strength in liquid water relative to PBE, a finding that is also in agreement with the previously observed improvements in the infrared spectrum of PBE0-based liquid water.~\cite{zhang_jctc_2011-2} 
In fact, this effect has also been demonstrated in previous studies on both gas-phase water clusters~\cite{santra_jcp_2009,gillan_jcp_2012,wang_jcp_2010} and ice,~\cite{santra_prl_2011,labat_jcc_2011,kambara_pccp_2012} in which a significant reduction in the strength of the hydrogen bonds was observed in each of these aqueous systems with the use of hybrid XC functionals.
By weakening the individual hydrogen bonds, simulations at the PBE0 level of theory yield a marked increase in the population of water molecules located in the interstitial region, \textit{i.e.}, the region between the first and second coordination shells.
However, this trend of destructuring or softening in the $g_{\rm OO}(r)$ of liquid water observed here is in contrast with a few previous studies at the PBE0 level of theory, in which no effect was observed on the $g_{\rm OO}(r)$ when exact exchange was accounted for over the standard PBE functional.~\cite{guidon_jcp_2008,ben_jpcl_2013}
In this regard, we note the existence of a theoretical difference in the treatment of the exact exchange contributions between this work, in which the exact exchange contributions were exactly computed numerically~\cite{wu_prb_2009} (See Sec.~\ref{sec:Theory}), and that of Guidon \textit{et al.},~\cite{guidon_jcp_2008} in which the Coulomb potential in the exact exchange integrals was replaced by a finite-range approximant in terms of the complementary error function, ${\rm erfc}(r)$.

By including a self-consistent treatment of the non-local vdW/dispersion interactions \textit{via} the density-dependent Tkatchenko-Scheffler dispersion correction,~\cite{tkatchenko_prl_2009} we observed a very similar effect on the $g_{\rm OO}(r)$ of liquid water; in fact, the radial distribution functions obtained at the PBE0 and PBE+vdW levels of theory are almost identical over the entire distance range considered (See Fig.~\ref{fig:gOO} and Table~\ref{tab1}).
Unlike exact exchange, however, vdW forces are non-directional and strengthen the interactions between a given water molecule and the water molecules in its first and second coordination shells, which actually causes the water molecules in the second coordination to move inward and populate the interstitial region.
As a result, the hydrogen bonding (See Sec.~\ref{sec:HB}) in the first coordination shell is weakened, causing the water molecules in the first coordination shell to be pushed outward as well. 
Here vdW forces tend to stabilize such disordered configurations---structures that would be energetically disfavored at the PBE level of theory---and this effect leads to the same net result in the $g_{\rm OO}(r)$ as that observed in PBE0-based liquid water.
In this regard, there is only a subtle difference in the $g_{\rm OO}(r)$ of liquid water between the PBE+vdW and PBE0 levels of theory: for PBE+vdW, the position of the first minimum, $r^{\rm min}_{1}$ is slightly decreased with respect to PBE by approximately 0.01 \AA\, as opposed to the increase of 0.02 \AA\, observed above for PBE0.
Due to this decrease in $r^{\rm min}_{1}$ at the PBE+vdW level of theory, the integrated number of water molecules in the first coordination shell,~\cite{coordination_number_footnote} $n_{\rm OO}$, actually decreases slightly when compared to PBE (4.01 vs. 4.03), whereas at the PBE0 level, $n_{\rm OO}$ increases to a value of 4.10 since $r^{\rm min}_{1}$ increases (See Table~\ref{tab1}). 

Several studies performed to date have concluded that the structure of liquid water significantly softens when vdW interactions are accounted for in AIMD simulations; however, the extent to which non-local vdW/dispersion forces affect the structure of liquid water is largely dependent on the given approach utilized to facilitate vdW-inclusive DFT.~\cite{lin_jpcb_2009,jonchiere_jcp_2011,wang_jcp_2011,zhang_jctc_2011,mogelhoj_jpcb_2011,lin_jctc_2012,yoo_jcp_2012,schmidt_jpcb_2009,ma_jcp_2012,ben_jpcl_2013}.
For instance, the most commonly utilized approach involves the pairwise-additive $C_6/R^6$ vdW/dispersion correction of Grimme~\cite{grimme_jcc_2004,grimme_jcp_2010} in conjunction with the popular PBE and BLYP GGA functionals, which tends to reduce the intensity of the first maximum in the $g_{\rm OO}(r)$ of liquid water over a fairly wide range, \textit{i.e.}, by approximately 5\% for PBE-D~\cite{schmidt_jpcb_2009} and 10-17\% for BLYP-D.~\cite{lin_jpcb_2009,schmidt_jpcb_2009,jonchiere_jcp_2011}
Other popular vdW-inclusive DFT approaches for studying liquid water include the so-called vdW-DF functionals,~\cite{dion_prl_2004} which incorporate vdW/dispersion interactions \textit{via} the use of explicit non-local correlation functionals.
This class of vdW-inclusive functionals have severe shortcomings in reproducing the second coordination shell in the $g_{\rm OO}(r)$, especially when employed in conjunction with the revPBE exchange functional;~\cite{wang_jcp_2011,mogelhoj_jpcb_2011} the use of the PBE exchange functional instead recovers this limitation to some extent~\cite{wang_jcp_2011,zhang_jctc_2011,corsetti_jcp_2013} and reduces the intensity of the first maximum by $\approx$ 25\% with respect to PBE.~\cite{zhang_jctc_2011}
For comparison, the self-consistent implementation of the density-dependent Tkatchenko-Scheffler~\cite{tkatchenko_prl_2009} vdW/dispersion correction employed in this work, PBE+vdW, yields an 8.8\% reduction in the intensity of the first maximum with respect to PBE.

The collective effect of treating exact exchange and non-local vdW/dispersion interactions, as depicted by the AIMD simulations of liquid water at the PBE0+vdW level of theory, yields a $g_{\rm OO}(r)$ that is even closer to the experimental findings.
In this case, both of these improvements to the underlying XC functional work together to reduce the intensity of the first maximum, $g^{\rm max}_{1}$, by 0.52 and to increase the intensity of the first minimum, $g^{\rm min}_{1}$, by 0.33, both of which represent larger effects than observed when the PBE0 and PBE+vdW XC functionals were utilized independently.
The collective reduction here in the first maximum of 0.52 at the PBE0+vdW level of theory is slightly less than the sum of the individual reductions observed when utilizing PBE0 (0.32) and PBE+vdW (0.29) independently, which is indicative of the fact that there is a compensating effect present here in the short-range (\textit{i.e.}, at distances which include hydrogen-bonded oxygens) when utilizing a self-consistent vdW/dispersion correction scheme that depends on the underlying charge density (\textit{i.e.}, the vdW/dispersion correction is a functional of the electron density, $E_{vdW}=E_{vdw}[n(\mathbf{r})]$).
In the same breath, the collective increase in the first minimum of 0.33 at the PBE0+vdW level of theory is exactly equal to the sum of the individual increases observed when utilizing PBE0 (0.16) and PBE+vdW (0.17) alone, reflecting the fact that both of these effects definitively work in unison to weaken the hydrogen bond network and increase the relative population of water molecules in the interstitial region.
We note in passing that similar conclusions can be drawn by considering the resultant shifts in the positions of the first maximum and minimum of the PBE0+vdW $g_{\rm OO}(r)$: the collective shift in the position of the first maximum of 0.03 \AA\, is only slightly smaller than the sum of the individual shifts of 0.02 \AA\, (PBE0) and 0.02 \AA\, (PBE+vdW), while the increase in the position of the first minimum of 0.03 \AA\, represents an interesting collective effect between PBE0 (which increased $r^{\rm min}_{1}$ by 0.02 \AA) and PBE+vdW (which decreased $r^{\rm min}_{1}$ by 0.01 \AA).
Furthermore, this outward shift in $r^{\rm max}_{1}$ at the PBE0+vdW level, while in the right direction, is still not large enough to exactly match the experimental findings, which is suggestive that a further increase in the amount of exact exchange may still be needed in the underlying XC potential.

Although a more detailed analysis of the nature and extent of the hydrogen bonding network generated from the AIMD simulations performed in this work will be discussed in Sec.~\ref{sec:HB}, another interesting point can be made here concerning the contributions arising from select neighboring water molecules to the overall $g(r)$ as a function of the underlying XC potential.
\begin{figure}
\begin{center}
\includegraphics[width=8.75cm]{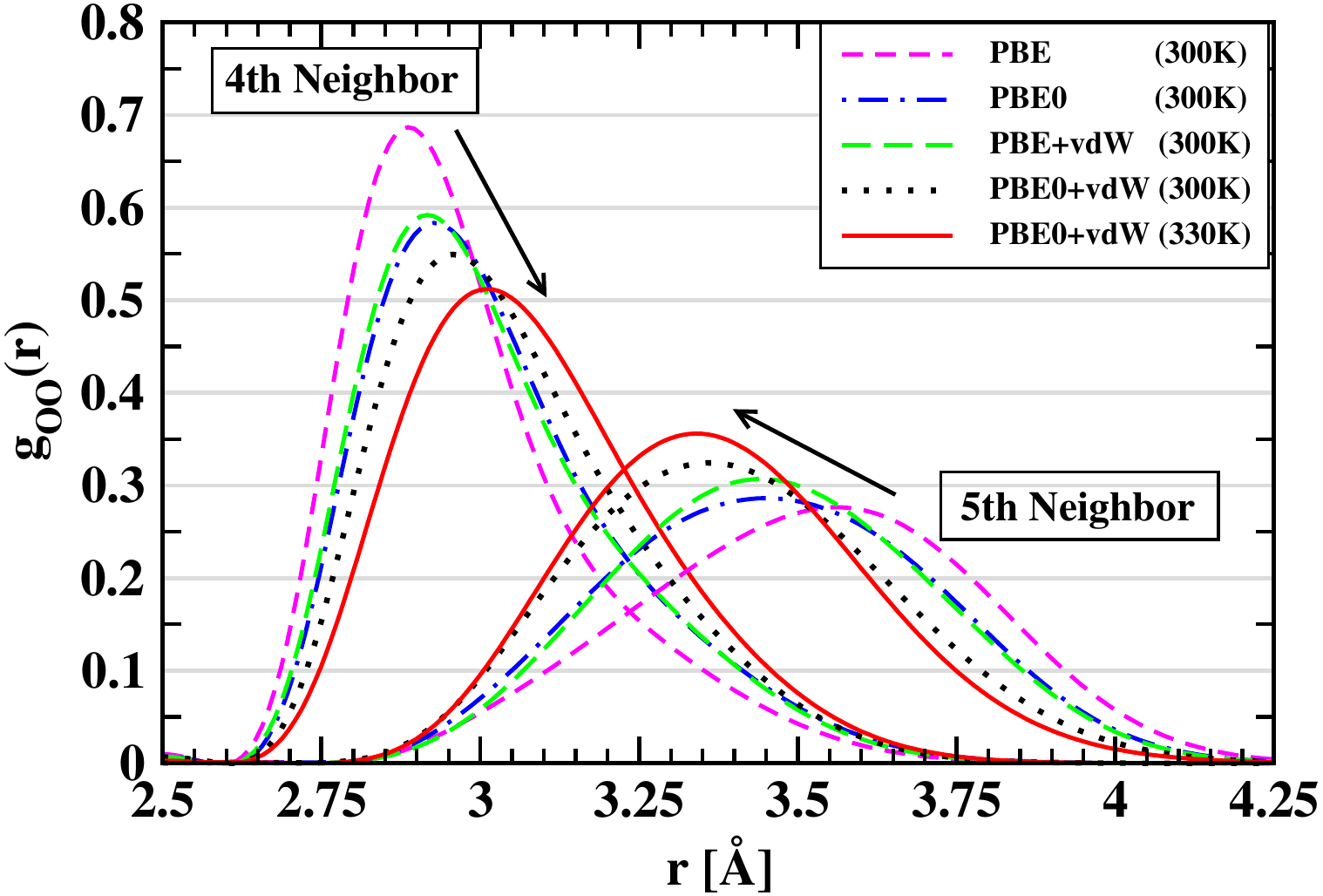}
\caption{\label{fig:gOOn} Contributions from the fourth (4th) and fifth (5th) neighbors (neighboring water molecules) to the oxygen-oxygen radial distribution functions $g_{\rm OO}(r)$ of liquid water obtained from theory (\textit{via} the DFT-based AIMD simulations performed in this work). The arrows indicate systematic shifts in the main peak positions and intensities of the computed distributions with improvement of the underlying exchange-correlation functional.}
\end{center}
\end{figure}
In Fig.~\ref{fig:gOOn}, the contributions to the $g_{\rm OO}(r)$ of liquid water arising from the fourth (4th) and fifth (5th) neighboring water molecules are plotted for each of the XC functionals employed in this work.
From this figure, one can immediately notice an apparent shift in the $g_{\rm OO}(r)$ contributions occurring between the 4th and 5th neighbors as the underlying XC functional is systematically improved by accounting for exact exchange and/or non-local vdW/dispersion interactions.
For the first four neighboring water molecules (with only the 4th neighbor shown in Fig.~\ref{fig:gOOn} for clarity), we observe that the individual and collective effects of Exx and vdW tend to simultaneously increase the position and decrease the intensity of this contribution to the overall $g_{\rm OO}(r)$; on the other hand, we observe the exact opposite for the fifth and sixth neighboring water molecules (with only the 5th neighbor shown in Fig.~\ref{fig:gOOn} for clarity), which tend to simultaneously decrease the position and increase the intensity of their contribution to the overall $g_{\rm OO}(r)$.
Both of these findings are indicative of a deviation from the perfect tetrahedral hydrogen bonding network observed in ice (and to a large extent in PBE liquid water) and a relative increase in the population of water molecules in the interstitial region.
As such, this clearly illustrates how these systematically improved XC functionals soften the structure of liquid water with respect to GGA-DFT and thereby generate oxygen-oxygen radial distribution functions that are increasingly in better agreement with the experimental data.

Even further improvements with respect to the experimental findings is possible when nuclear quantum effects are mimicked in the $g_{\rm OO}(r)$ of liquid water by performing the simulation at the PBE0+vdW level of theory with a 30 K increase in the simulation temperature (See Fig.~\ref{fig:gOO} and Table~\ref{tab1}).
At 330 K, we observe further softening in the $g_{\rm OO}(r)$ of liquid water, with peak positions and intensities that are now in quantitative agreement with the $g_{\rm OO}(r)$ directly obtained from X-ray~\cite{skinner_jcp_2013} scattering experiments.
In this regard, approximately accounting for NQE leads to a further decrease in the intensity of the first maximum of 0.25, which is now approximately 2.3-4.2\% lower than the experimental data, and a further increase in the intensity of the first minimum of 0.14, which is in exact agreement with Skinner \textit{et al.}~\cite{skinner_jcp_2013} and only 2.4\% higher than the value provided by Soper and Benmore.~\cite{soper_prl_2008}
Similar agreement can be found when analyzing the positions of the first maximum (with an error of 0.7-2.1\%) and the first minimum (with an error of 2.6-3.5\%) with respect to the experimental findings, as well as the resulting coordination number. 
This level of agreement with experiment in the $g_{\rm OO}(r)$ of liquid water has been quite difficult to achieve to date and clearly requires an increase in the accuracy of the underlying XC functional, and at the same time, suggests the need for a quantum mechanical treatment of the nuclear degrees of freedom.

\subsection{Comparative Analysis of the Oxygen-Hydrogen Radial Distribution Function \label{sec:OHRDF}}

\begin{figure}
\begin{center}
\includegraphics[width=8.75cm]{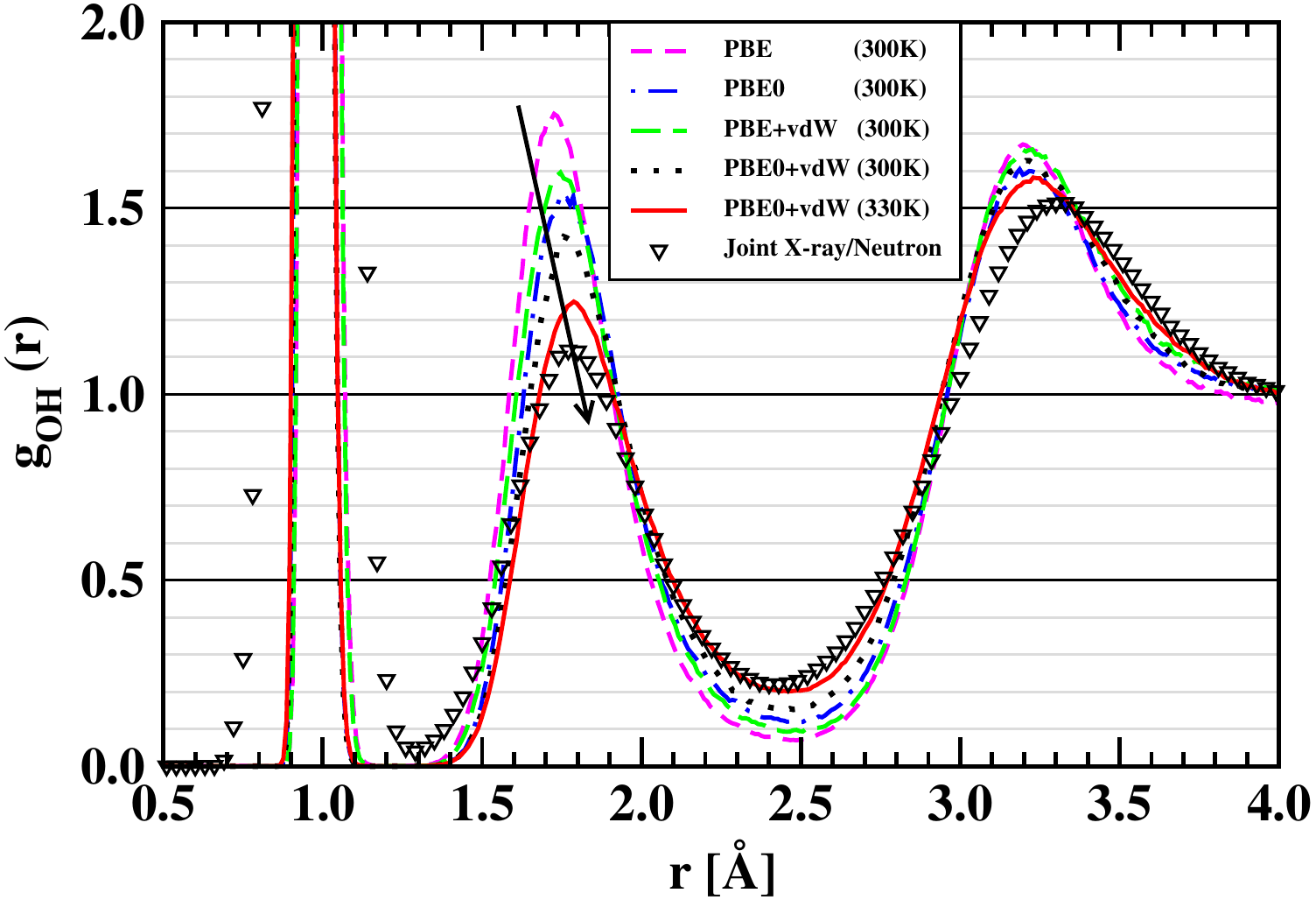}
\caption{\label{fig:gOH} The oxygen-hydrogen radial distribution functions $g_{\rm OH}(r)$ of liquid water obtained from theory (\textit{via} the DFT-based AIMD simulations performed in this work) and scattering experiments.~\cite{soper_prl_2008} The arrow indicates systematic shifts in the main peak positions and intensities of the computed distributions with improvement of the underlying exchange-correlation functional.}
\end{center}
\end{figure}

A similar trend of systematic improvements in the theoretical description of the oxygen-hydrogen radial distribution function, $g_{\rm OH}(r)$, of liquid water (See Fig.~\ref{fig:gOH}) can be observed with the inclusion of exact exchange (\textit{via} the PBE0 hybrid functional), non-local vdW/dispersion interactions (\textit{via} a self-consistent density-dependent $C_6/R^6$ correction), and approximate nuclear quantum effects (\textit{via} a 30 K increase in the simulation temperature).
In this case, however, the width and intensity of the first peak of the $g_{\rm OH}(r)$ of liquid water, which describes the covalent O-H bond, can only be described by properly accounting for nuclear quantum effects in the simulation of liquid water, \textit{i.e.}, the approximate treatment of nuclear quantum effects \textit{via} a 30 K increase in the simulation temperature is not sufficient to capture the quantum nature of the lighter hydrogen atoms, which deviate significantly from classical behavior at room temperature.~\cite{morrone_prl_2008,ceriotti_pnas_2013}
As such, the explicit inclusion of NQE in the \textit{ab initio} simulations of liquid water \textit{via} the Feynman discretized path integral (PI) scheme is the focus of intense ongoing research in our group and will be addressed in future work.

The largest effects arising from the individual and collective treatment of exact exchange and non-local vdW/dispersion interactions can be found in the position and intensity of the second maximum in the $g_{\rm OH}(r)$, which is the signature of the hydrogen bond network in liquid water.
In this regard, we observe very similar trends to those reported above for the $g_{\rm OO}(r)$ of liquid water, in that the AIMD simulations at the PBE0 and PBE+vdW levels of theory predict nearly identical $g_{\rm OH}(r)$ over the entire distance range considered (See Fig.~\ref{fig:gOH}).
Considering for instance the intensity of the second maximum in the $g_{\rm OH}(r)$, there is an overall reduction or destructuring of approximately 0.22 (PBE0) and 0.15 (PBE+vdW), which is consistent with the fact that both of these XC functionals tend to disrupt the hydrogen bond network and increase the relative population of water molecules in the interstitial region.
In this case, the effect of including a fraction of exact exchange on the $g_{\rm OH}(r)$ is only slightly larger than the observed effect resulting from an explicit treatment of vdW/dispersion interactions.
When used in combination, the PBE0+vdW XC functional again displays a nearly additive effect on the $g_{\rm OH}(r)$ by reducing the intensity of the second maximum by 0.33, which is to be contrasted against the sum of the aforementioned individual contributions at 0.37.

An even further reduction in the intensity of the second maximum in the $g_{\rm OH}(r)$ is provided by PBE0+vdW AIMD simulations performed at 330 K, which yield an overestimation of approximately 11.6\% with respect to the joint X-ray/neutron scattering results of Soper and Benmore.~\cite{soper_prl_2008}
We again stress here that structural quantities that directly involve the hydrogen atoms in liquid water, such as the $g_{\rm OH}(r)$ and $g_{\rm HH}(r)$, require an explicit treatment of nuclear quantum effects; hence, errors in excess of 10\% as reported above are completely reasonable and expected in such quantities when generated from simulations employing classical mechanics for the nuclear equations of motion.
In the same breath, structural quantities like the $g_{\rm OO}(r)$, which was considered in detail in Sec.~\ref{sec:OORDF}, were less sensitive to an approximate treatment of NQE, and nearly quantitative accuracy can be achieved for this property with a 30 K increase in the simulation temperature, provided a sufficiently accurate underlying XC potential is utilized.

\subsection{Tetrahedrality and the Oxygen-Oxygen-Oxygen Triplet Distribution Function \label{sec:OOO}}

\begin{figure}
\begin{center}
\includegraphics[width=8.75cm]{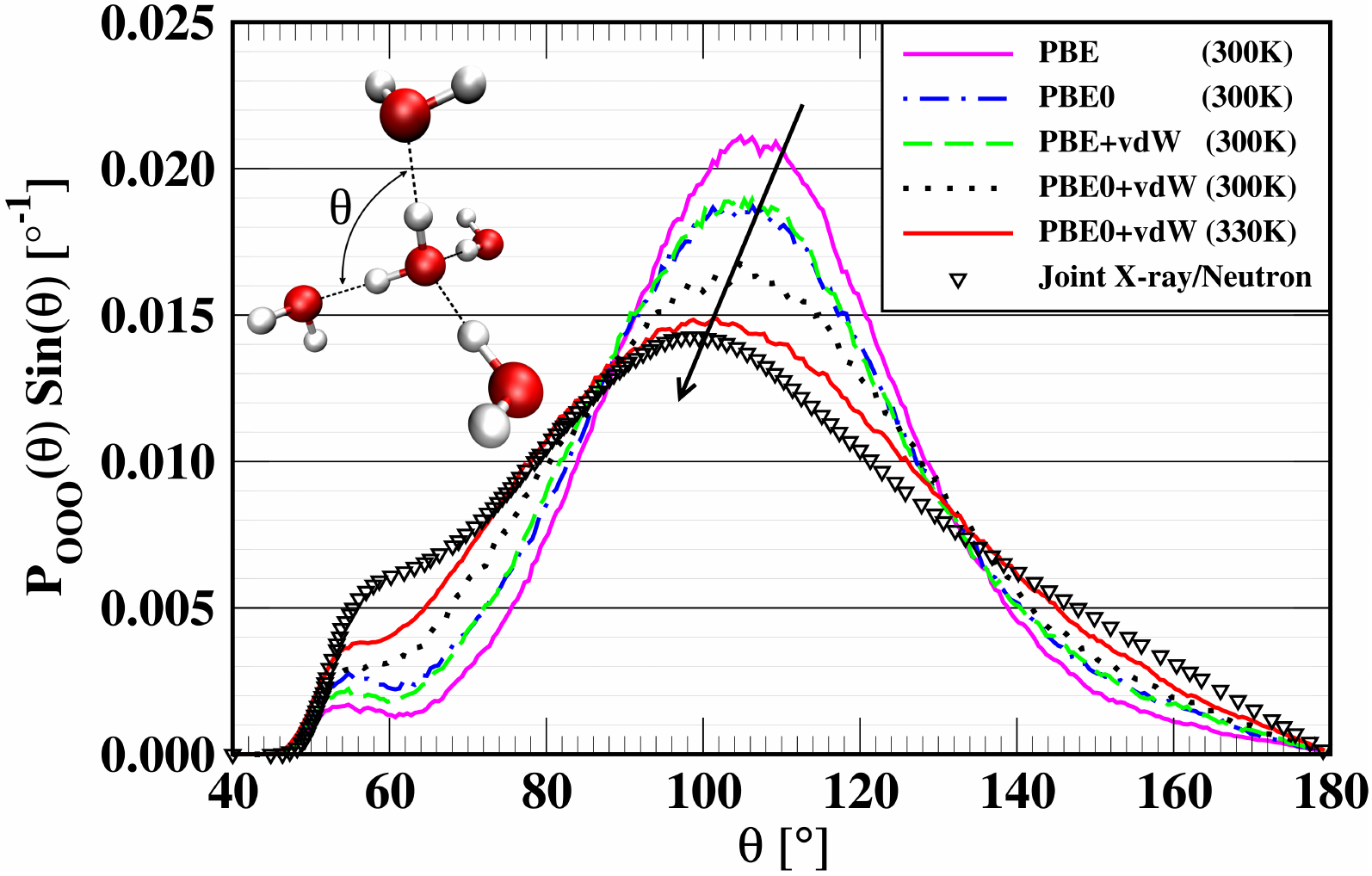}
\caption{\label{fig:pOOO} The oxygen-oxygen-oxygen triplet angular distribution functions $P_{\rm OOO}(\theta)$ of liquid water obtained from theory (\textit{via} the DFT-based AIMD simulations performed in this work) and empirical potential structural refinement (EPSR) based on joint X-ray/neutron scattering data.~\cite{soper_prl_2008} The triplet angular distribution functions show here were normalized to $\int_0^{\pi} d\theta\,P_{\rm OOO}(\theta){\rm sin}(\theta)=1$. The arrow indicates systematic shifts in the main peak positions and intensities of the computed distributions with improvement of the underlying exchange-correlation functional.}
\end{center}
\end{figure}

In order to further analyze the local arrangement of water molecules in condensed-phase liquid water, we have computed the distribution of triplet oxygen-oxygen-oxygen angles, $P_{\rm OOO}(\theta)$, within the first coordination shell (See Fig.~\ref{fig:pOOO}).
In general, three-body correlation functions such as $P_{\rm OOO}(\theta)$ are not directly accessible from scattering experiments; however, an experimental distribution of $\theta$ was recently extracted \textit{via} empirical potential structural refinement (EPSR) based on joint X-ray/neutron scattering data.~\cite{soper_prl_2008}
In this regard, we note that the aforementioned quantitative agreement between the oxygen-oxygen radial distribution functions (See Fig.~\ref{fig:gOO}) directly obtained \textit{via} X-ray scattering experiments~\cite{skinner_jcp_2013} and EPSR based on joint X-ray/neutron scattering data~\cite{soper_prl_2008} supports the effectiveness of this procedure.
To extract the corresponding $P_{\rm OOO}(\theta)$ quantity from the AIMD simulations performed in this work, three oxygen atoms were considered as part of a given triplet if two of the oxygen atoms were within a prescribed cutoff distance from the third. 
Following Ref.~[\onlinecite{soper_prl_2008}], this cutoff distance was chosen to yield an average oxygen-oxygen coordination number of $\approx$ 4, and fell within the range of 3.25-3.27 \AA$\,$ for all of the AIMD simulations considered herein.

From Fig.~\ref{fig:pOOO}, the EPSR-experimental $P_{\rm OOO}(\theta)$ shows a broad peak around 100$^\circ$, which is indicative of the fact that the local tetrahedral network in liquid water is substantially more disordered than in crystalline ice.
At the PBE-GGA level of theory, we find that the peak of $P_{\rm OOO}(\theta)$ is very close to the perfect tetrahedral angle of 109.5$^\circ$ and the overall distribution is much too narrow as compared to the EPSR-experimental results---another manifestation of the fact that liquid water generated using PBE is overstructured and the degree of local tetrahedral order is much higher than that observed in experiment.
In fact, by computing the tetrahedral order parameter, $q$, as~\cite{errington_nature_2001}
\begin{equation}
\label{tetparam}
q=1-\frac{3}{8} \sum_{i=1}^3 \sum_{j=i+1}^4 \left({\rm cos}(\theta_{ij})+\frac{1}{3}\right)^2 ,
\end{equation}
in which $\theta_{ij}$ is the angle formed by a central given water molecule and its nearest neighbors $i$ and $j$, we find that PBE yields an average $q$ value of 0.78, which is indeed much higher than the estimated EPSR-experimental value of $q=0.576$~\cite{soper_prl_2008} (See Table~\ref{tab1}).
For reference, a system with perfect tetrahedral order would yield $q=1$, whereas $q=0$ for the ideal gas (\textit{i.e.}, a system with random positions).

As seen above, the degree of local tetrahedral order in liquid water can be significantly reduced when utilizing an XC potential that accounts for exact exchange and non-local vdW/dispersion interactions.
As shown in Fig.~\ref{fig:pOOO}, the individual and collective effects of Exx and vdW shift the peak of $P_{\rm OOO}(\theta)$ towards lower $\theta$ values and make the overall distribution much broader, both of which result in better agreement with the EPSR-experimental triplet distribution functions.
Here we note that a reduction in the tetrahedral order parameter with respect to the GGA-DFT values has been previously reported using other vdW-inclusive functionals~\cite{jonchiere_jcp_2011,zhang_jpcb_2011} with values of $q\approx0.7$, which is in reasonable agreement with the value of $q=0.74$ computed herein at the PBE+vdW level of theory; however, capturing this trend using several vdW-inclusive functionals has come with the deleterious cost of a nearly vanishing second coordination shell in the $g_{\rm OO}(r)$.~\cite{mogelhoj_jpcb_2011} 
At the PBE0 level, we computed a value of $q=0.73$ for the tetrahedral order parameter, which is extremely close to the value obtained above in the PBE+vdW case and in good agreement with the previous study of Zhang \textit{et al.}~\cite{zhang_jpcb_2011} 
When used in conjunction, the self-consistent PBE0+vdW exchange-correlation functional yields a value of $q=0.69$ for the tetrahedral order parameter, which is exactly additive in terms of the aforementioned individual effects of Exx and vdW and therefore closer to the EPSR-experimental measure of the local tetrahedrality.

As seen above in Sec.~\ref{sec:OOSF}--\ref{sec:OHRDF}, even further improvement comes with the approximate inclusion of NQE, wherein we computed a value of $q=0.64$, which is now approximately 10.3\% larger than the EPSR-experimental value.
Interestingly, the shoulder present in the EPSR-experimental $P_{\rm OOO}(\theta)$ distribution at approximately $60^\circ$, which is indicative of a highly distorted hydrogen bond network in the first coordination shell, only becomes noticeably visible at the PBE0+vdW (330 K) level of theory.
Nevertheless, this quantity is sensitive to the positions of the hydrogen atoms, and as such, will require a proper treatment of nuclear quantum effects in order to accurately capture these finer details.

\subsection{Hydrogen Bond Analysis in Liquid Water \label{sec:HB}}

The fluidity in liquid water comes from the fact that the hydrogen bonds, which create the underlying random tetrahedral network, are continuously breaking and forming.
Unlike ice, in which each water molecule is involved in four hydrogen bonds with its nearest neighbors (\textit{i.e.}, each water molecule is simultaneously donating and accepting two hydrogen bonds), liquid water contains some fraction of broken hydrogen bonds on average.~\cite{eaves_pnas_2005}
In this regard, any quantitative measure of the number of either intact or broken hydrogen bonds in liquid water is somewhat ambiguous, since the notion of a hydrogen bond itself is not uniquely defined.
However, qualitative agreement between many proposed definitions for intact hydrogen bonds (\textit{i.e.}, those based on geometry, topology, and even electronic structure) have been deemed satisfactory over a wide range of thermodynamic conditions.~\cite{prada-gracia_jcp_2013}
Since we are interested in the qualitative change in the statistics of the number and types of hydrogen bonds in liquid water as a function of the individual and collective treatment of exact exchange, non-local vdW/dispersion interactions, and approximate nuclear quantum effects, we have chosen to utilize the popular hydrogen bond definition proposed by Luzar and Chandler,~\cite{luzar_prl_2006} wherein the defining parameters for an intact hydrogen bond include both a distance and angular criterion, namely, $R_{\rm OO}<3.5$ \AA\, and $\beta<30^\circ$, with $\beta\equiv\angle {\rm O_A \cdots O_D\!\!-\!\!H_D}$.

Using this definition, we computed the average number of intact hydrogen bonds per water molecule~\cite{hbond_footnote} for each of the AIMD simulation performed in this work and reported these numbers in Table~\ref{tab1}.
From this data, we first observed that the average number of intact hydrogen bonds per water molecule decreases in the following order as the underlying XC potential is systematically improved: PBE (3.83), PBE+vdW (3.74), PBE0 (3.71), PBE0+vdW (3.67), and PBE0+vdW+aNQE (3.62). 
This decrease in the number of intact hydrogen bonds per water molecule (or corresponding increase in the number of broken hydrogen bonds per water molecule) is representative of a larger degree of disorder in the local tetrahedral network in liquid water, and allows for an increase in the relative population of water molecules in the interstitial region, a feature that has already manifested itself in several of the structural quantities considered above, in particular in the reduction of the intensity of the first minimum in the $g_{\rm OO}(r)$ in Fig.~\ref{fig:gOO}.

\begin{figure}
\begin{center}
\includegraphics[width=8.75cm]{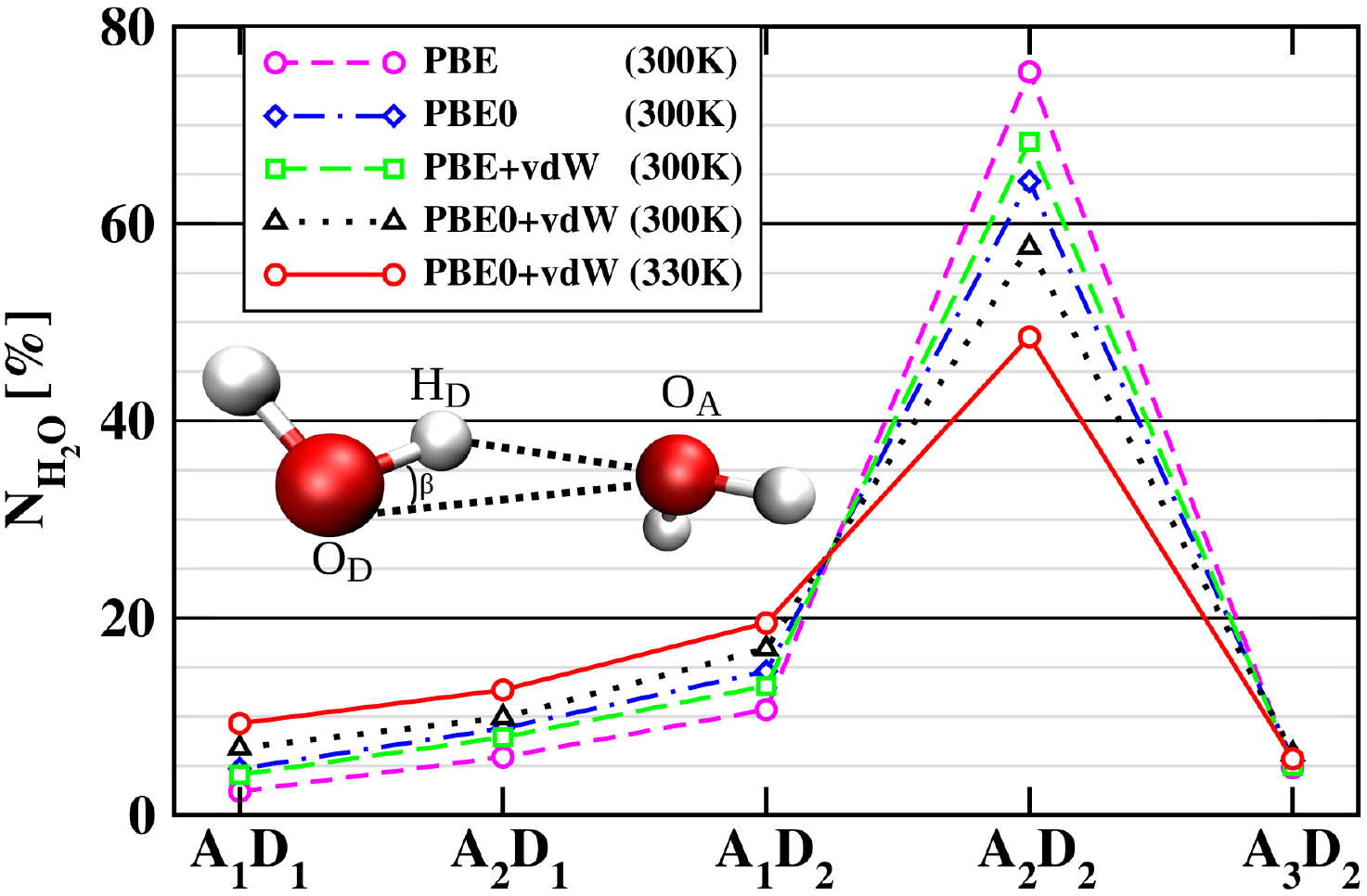}
\caption{\label{fig:hbda} Percentage-wise decomposition into acceptor- (A) and donor- (D) types of the intact hydrogen bonds per water molecule in liquid water obtained from theory (\textit{via} the DFT-based AIMD simulations performed in this work). The $x$-axis labels ${\rm A}_x{\rm D}_y$ indicate the number of acceptor-type (${\rm A}_x$) and donor-type (${\rm D}_y$) hydrogen bonds, respectively. All other acceptor--donor combinations, \textit{e.g.}, ${\rm A}_3{\rm D}_1$, ${\rm A}_0{\rm D}_y$, ${\rm A}_x{\rm D}_0$, etc., had contributions of $<1$\% and were therefore omitted from this figure for clarity. The inset depicts donor (O$_{\rm D}$) and acceptor (O$_{\rm A}$) oxygen atoms participating in a hydrogen bond via the hydrogen (H$_{\rm D}$) atom.}
\end{center}
\end{figure}

To provide a more detailed look into the hydrogen bonding in liquid water, we also performed a decomposition of the intact hydrogen bonds per water molecule into acceptor- (A) and donor- (D) types, \textit{i.e.}, ${\rm O_A \cdots H_D\!\!-\!\!O_D}$, in Fig.~\ref{fig:hbda}.
From this figure, one immediately notices that the largest percentage of the intact hydrogen bonds for each of the AIMD simulations considered herein are of the ${\rm A}_2 {\rm D}_2$ type---the hydrogen bonding motif which represents the traditional picture of a water molecule accepting and donating two hydrogen bonds.
At the PBE level of theory, the percentage of ${\rm A}_2 {\rm D}_2$ type hydrogen bonds is a majority at 75.4\%, a quantity which is reduced to just less than 50\% when Exx, vdW, and aNQE effects are accounted for in the underlying XC potential at the PBE0+vdW (330 K) level, a trend that again reflects the increasing level of disorder in the local tetrahedral network in liquid water.

Hydrogen bond types in which a given water molecule is involved in three hydrogen bonds, as in the ${\rm A}_1 {\rm D}_2$ and ${\rm A}_2 {\rm D}_1$ types, comprise the next largest percentage of the intact hydrogen bonds in liquid water.
Here we find an accompanying increase in the relative population of these hydrogen bond types as the underlying XC potential is improved from PBE to PBE0+vdW (330 K); in this case, the percentage of the ${\rm A}_1 {\rm D}_2$ and ${\rm A}_2 {\rm D}_1$ hydrogen bond types were observed to approximately double from 10.7\% to 19.5\% and 5.9\% to 12.7\%, respectively.
Expectedly, the number of intact hydrogen bonds in which a given water molecule is involved in less than three hydrogen bonds, as in the ${\rm A}_1 {\rm D}_1$ type, or more than four hydrogen bonds, as in the ${\rm A}_3 {\rm D}_2$ type, are less likely to occur and this trend is also reflected in Fig.~\ref{fig:hbda}. 
In this regard, we observed that the percentage of the ${\rm A}_1 {\rm D}_1$ hydrogen bond type approximately increased by a factor of four from 2.4\% (PBE) to 9.3\% (PBE0+vdW (330 K)), while the percentage of the ${\rm A}_3 {\rm D}_2$ type remained essentially constant and therefore independent of the underlying XC potential.

This decomposition analysis of the intact hydrogen bonds as a function of the underlying XC potential again reflects the individual and collective effects of Exx, vdW, and aNQE in the microscopic structure of liquid water; with Exx and aNQE weakening the hydrogen bond strength and non-local vdW/dispersion interactions stabilizing disordered water configurations, there is a relative increase in the population of interstitial water molecules which serves to introduce an increasing degree of disorder into the underlying tetrahedral network.

\subsection{Dipole Moment Analysis in Liquid Water \label{sec:DM}}

With the inclusion of exact exchange, non-local vdW/dispersion interactions, and approximate nuclear quantum effects, the modifications to the local microscopic structure of liquid water observed in this work can also be correlated with molecular electrostatic properties which govern the strength of the hydrogen bonding network and influence solvation behavior.
In this section, we investigate one such electrostatic property, namely the distribution of molecular dipole moments---the first non-zero multipole moments in the individual water molecules comprising the condensed phase.
For reference, the dipole moment of a single water molecule in the gas phase is accurately known to be 1.855 Debye (D) from spectroscopic measurements,~\cite{shostak_jcp_1991,clough_jcp_1973} and several DFT functionals can reproduce this experimental value to within 3\%.~\cite{zhang_jctc_2011}
However, there still remains a large uncertainty in the value of the molecular dipole moment in liquid water as extracted from X-ray scattering form factors ($2.9 \pm 0.6$ D),~\cite{badyal_jcp_2000} and in this regard DFT-based AIMD has the potential to provide molecular dipole moments in the condensed phase with a larger degree of certainty.

The calculation of molecular dipole moments in condensed-phase systems requires partitioning of the electron density, which again cannot be accomplished in a uniquely defined manner; as a result, the magnitude of the molecular dipole moment in ice Ih, for instance, can vary between 2.3--3.1 D depending on the partitioning scheme employed.~\cite{batista_jcp_1999}
In this regard, maximally localized Wannier functions (MLWFs), which are obtained \textit{via} a unitary transformation of the occupied Kohn-Sham electronic states, have been shown to be an extremely useful tool in computing molecular dipole moments in condensed-phase environments.~\cite{silvestrelli_jcp_1999,silvestrelli_prl_1999,sharma_prl_2005,zhang_jctc_2011,sharma_prl_2007}
For the case of liquid water, the MLWFs associated with different molecules have only a very minor overlap,~\cite{silvestrelli_jcp_1999} and this fact allows for a nearly unique definition of the charges belonging to a given water molecule, thereby eliminating to a large extent the aforementioned partitioning issues when computing molecular dipole moments in the condensed-phase.~\cite{sharma_prl_2005,sharma_prl_2007}
%
\begin{figure*}
\begin{center}
\includegraphics[width=17.5cm]{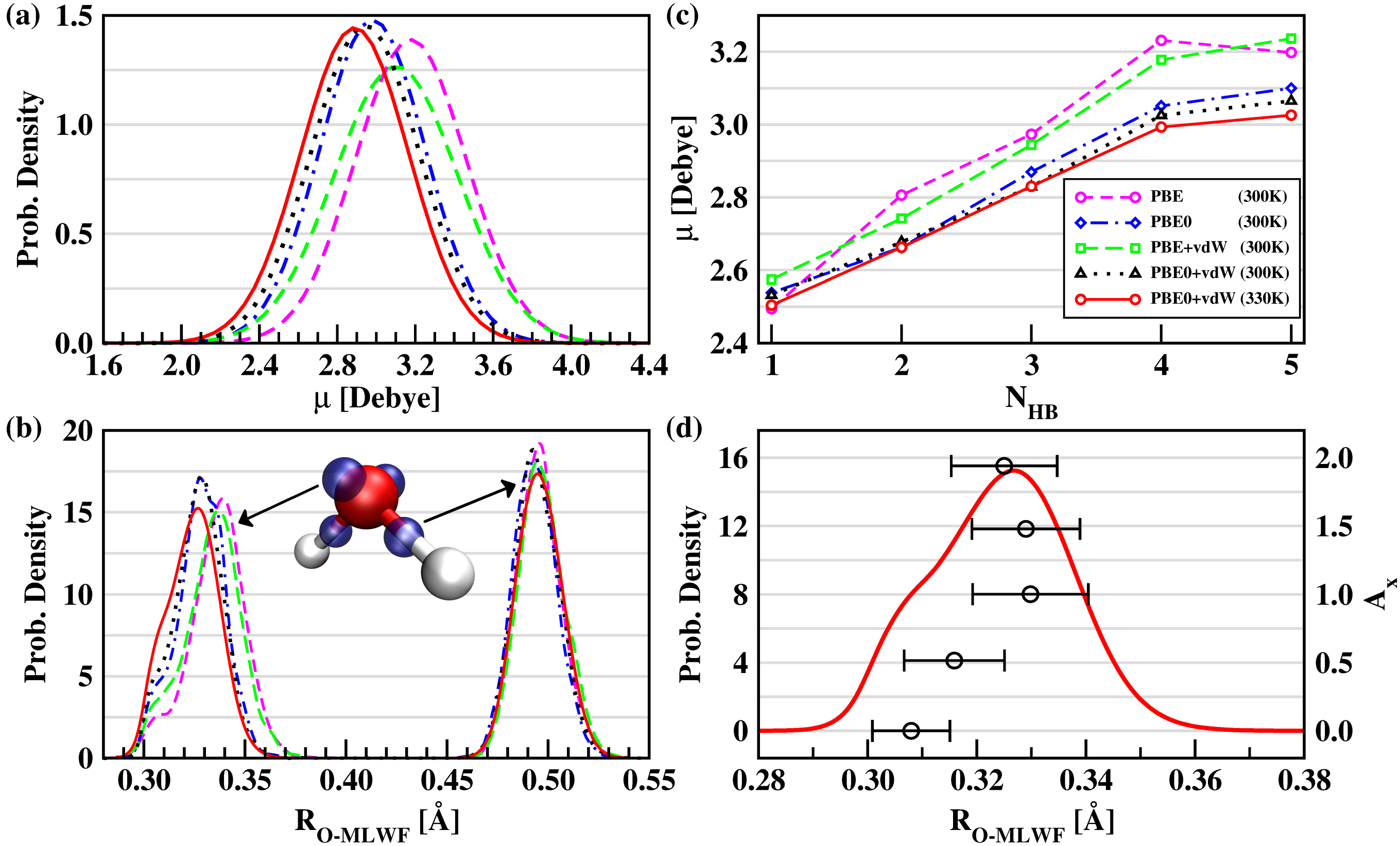}
\caption{\label{fig:dm} \textbf{Panel (a)}: Probability density plots of the distributions of molecular dipole moment magnitudes ($\mu$ in Debye) in liquid water obtained from theory (\textit{via} the DFT-based AIMD simulations performed in this work). \textbf{Panel (b)}: Probability density plots of the distributions of distances between the oxygen atoms in a given water molecule and the centers of the corresponding maximally localized Wannier functions ($R_{\rm O-MLWF}$ in \AA). \textbf{Panel (c)}: Plots of the variation in the magnitude of the molecular dipole moments ($\mu$ in Debye) with the number of intact hydrogen bonds ($N_{\rm HB}$) per water molecule. \textbf{Panel (d)}: Probability density plot of the O--MLWF distance distribution (\textit{red line, left axis}) reproduced from \textbf{Panel (b)} and the average number of acceptor-type (${\rm A}_x$) hydrogen bonds per water molecule (\textit{black circles, right axis}) as a function of $R_{\rm O-MLWF}$ in liquid water obtained from an AIMD simulation at the PBE0+vdW (330 K) level of theory.}
\end{center}
\end{figure*}

Using the four MLWFs corresponding to each water molecule (which represent the eight associated valence electrons), the molecular dipole moment ($\boldsymbol \mu$) for a water molecule in the condensed-phase can then be computed as
\begin{equation}
\boldsymbol \mu = \mathbf R_{\rm H_1} + \mathbf R_{\rm H_2} + 6\mathbf R_{\rm O} - 2\sum_{i=1}^{4} \mathbf R_{{\rm W}_i} ,
\end{equation}
in which $\mathbf R_{\rm H_1}$, $\mathbf R_{\rm H_2}$, and $\mathbf R_{\rm O}$ are the Cartesian coordinates of the hydrogen and oxygen atoms comprising the water molecule, respectively, and $\mathbf R_{{\rm W}_i}$ are the coordinates of the four corresponding MLWF centers.

Using this prescription, we have computed the distribution of molecular dipole moments in liquid water for each of the DFT-based AIMD simulations performed in this work and plotted the resulting data in Fig.~\ref{fig:dm}(a).
From this figure, it is evident that the peak positions of the dipole moment distributions are shifted towards smaller values as Exx, vdW, and aNQE effects are accounted for in the underlying XC potential.
In fact, the average magnitude of the molecular dipole moment in liquid water decreases in the following order as the underlying XC potential is systematically improved: PBE ($3.19\pm0.29$ D), PBE+vdW ($3.12\pm0.31$ D), PBE0 ($3.00\pm0.27$ D), PBE0+vdW ($2.96\pm0.27$ D), and PBE0+vdW+aNQE ($2.91\pm0.28$ D).
Unlike the structural properties considered above, we note that the inclusion of exact exchange in the XC potential clearly has a larger effect on the magnitude of the molecular dipole moment in liquid water than the albeit self-consistent treatment of non-local vdW/dispersion interactions employed herein.
Since the dipole moment is an electrostatic property, this result is not surprising as the hybrid PBE0 functional induces changes in the underlying local electronic structure of the individual water molecules that comprise the condensed-phase.~\cite{zhang_jctc_2011}
Although each of the aforementioned functionals yields a molecular dipole moment magnitude that is within the (relatively large) error bar of the experimental data, the value of $2.91\pm0.28$ D obtained at the PBE0+vdW (330 K) level of theory is very close to the mean experimental value and is likely to be the most accurate estimate of this quantity among the series of XC functionals considered herein.

To further investigate the variation in the molecular dipole moment in liquid water as a function of the underlying XC potential, we also computed the distribution of distances between the oxygen atoms in a given water molecule and the centers of the corresponding MLWFs ($R_{\rm O-MLWF} = |\mathbf{R}_{\rm O}-\mathbf{R}_{\rm W}|$) for each AIMD simulation performed in this work (See Fig.~\ref{fig:dm}(b)).
In a given water molecule, the corresponding set of four associated MLWFs can straightforwardly be identified as (\textit{i}) two bonding electron pairs, centered along the O-H covalent bonds with $R_{\rm O-MLWF} \approx 0.5$ \AA\, from the central oxygen atom and (\textit{ii}) two lone electron pairs, oriented in an essentially tetrahedral fashion with respect to the bonding electron pairs, but centered at significantly shorter distances from the central oxygen atom ($R_{\rm O-MLWF} = 0.30-0.35$ \AA).
From Fig.~\ref{fig:dm}(b), it is clear that the distribution of bonding pair O-MLWF distances is essentially independent of the underlying XC potential, whereas the peak positions of the lone pair O-MLWF distance distributions systematically shift toward shorter distances (by approximately $0.01-0.02$ \AA) as Exx, vdW, and aNQE effects are accounted for in the AIMD simulations.
This net reduction in the lone pair O-MLWF distance is indicative of a decrease in the magnitude of the local molecular dipole moment and therefore a weakened hydrogen bond network in liquid water, since a given water molecule accepts a hydrogen bond \textit{via} the electrostatic interaction between one of its lone electron pairs (represented here by the MLWF) and a hydrogen atom situated on the corresponding donor water molecule.
In fact, this correlation is immediately visible from the plot of the variation in the magnitude of the molecular dipole moments with the number of intact hydrogen bonds ($N_{\rm HB}$) per water molecule as a function of the underlying XC potential in Fig.~\ref{fig:dm}(c).
From this data, one immediately notices an appreciably direct (and nearly linear) correlation between the number of intact hydrogen bonds per water molecule and the strength of the molecular dipole moment.
Hence the systematic reduction in the strength and extent of the hydrogen bonding network as a function of systematically improving the underlying XC potential as discussed above is intimately related to the reduction in the magnitude of the local molecular dipole moment in liquid water.
This trend was observed with all of the functionals considered herein and is consistent with the systematic decrease in the population of water molecules having four intact hydrogen bonds (and the corresponding increase in the population of water molecules with two and three intact hydrogen bonds) as illustrated in Sec.~\ref{sec:HB} and previous studies on gas-phase water clusters.~\cite{batista_jcp_1999,kemp_jpca_2008,tu_cpl_2000}

One can take this analysis one step further by investigating the correlation between the small shoulder situated at $\approx$ 0.31 \AA\, in the lone pair O-MLWF distance distribution---a feature which becomes increasingly more pronounced with the collective inclusion of Exx, vdW, and aNQE in the underlying XC potential (See Fig.~\ref{fig:dm}(b))---and the number of intact acceptor-type hydrogen bonds per water molecule.
In Fig.~\ref{fig:dm}(d), we performed such an analysis by assigning~\cite{mlwf_footnote} the intact acceptor-type hydrogen bonds to either of the two MLWFs associated with the lone electron pairs of the corresponding acceptor water molecule for the AIMD simulation at the PBE0+vdW (330 K) level of theory.
In this regard, we observed that the number of acceptor-type hydrogen bonds decreases as the lone pair O-MLWF distance becomes shorter, a result that is again consistent with our above findings and indicative of a strong correlation between the underlying electronic structure of a given water molecule and the overall strength and integrity of the hydrogen bond network in liquid water.

\section{Conclusions \label{sec:Conclusions}}

In this work, we have performed a series of DFT-based AIMD simulations of ambient liquid water using a hierarchy of XC functionals to investigate the individual and collective effects of exact exchange (Exx), non-local vdW/dispersion interactions, and an approximate treatment of nuclear quantum effects (aNQE) on the microscopic structure of liquid water.
Utilizing highly efficient linear scaling algorithms as developed and extensively optimized in our group,~\cite{wu_prb_2009} we have employed the PBE0 hybrid functional to account for Exx effects in conjunction with a fully self-consistent implementation of the charge density-dependent dispersion correction of Tkatchenko and Scheffler~\cite{tkatchenko_prl_2009} to account for vdW interactions. 
Based on these AIMD simulations, we found that the inclusion of Exx and vdW systematically reduces the overstructuring in structural quantities such as the oxygen-oxygen ($g_{\rm OO}(r)$) and oxygen-hydrogen ($g_{\rm OH}(r)$) radial distribution functions that is commonly observed in liquid water generated by GGA-DFT based AIMD simulations.
Moreover, the collective effects of Exx, vdW, and aNQE, as resulting from a large-scale AIMD simulation of (H$_2$O)$_{128}$ at the PBE0+vdW level of theory (coupled with a 30 K increase in the simulation temperature to mimic the NQE) yield an oxygen-oxygen structure factor, $S_{\rm OO}(Q)$, and corresponding $g_{\rm OO}(r)$ that are in quantitative agreement with the best available experimental data.
The molecular conformations obtained from the AIMD simulations performed in this work depend on the microscopic interactions and the sampling methodology utilized herein.
Thus, the details of the simulated structures may change reflecting improvements in the underlying XC functional, stricter criteria for adiabatic separation of the nuclear and electronic coordinates, as well as an explicit quantum mechanical treatment of the nuclear degrees of freedom.
While these factors may affect the outcome of future simulations of liquid water, the qualitative effects found here should be robust and include: an increase in the relative population of water molecules in the interstitial region between the first and second coordination shells, a collective reorganization in the liquid phase which is facilitated by a weakening of the hydrogen bond strength by the use of the PBE0 hybrid XC functional, coupled with a relative stabilization of the resultant disordered liquid water configurations (\textit{i.e.}, configurations which significantly deviate from the perfect tetrahedral network in crystalline ice) by the inclusion of non-local vdW/dispersion interactions.
This increasingly more accurate description of the underlying hydrogen bond network in liquid water obtained at the PBE0+vdW+aNQE level of theory also yields higher-order correlation functions, such as the oxygen-oxygen-oxygen triplet angular distribution, $P_{\rm OOO}(\theta)$, and therefore the degree of local tetrahedrality, as well as electrostatic properties, such as the effective molecular dipole moment, that are in much better agreement with experiment.
As such, future research directions along this line include an investigation of the individual and collective effects of Exx, vdW, and NQE on the local environment and equilibrium density of ambient liquid water---additional structural properties that should be heavily influenced by an improved underlying description of the microscopic structure of liquid water. 
Furthermore, the accurate description of the underlying hydrogen bond network in liquid water---as provided by the theoretical methodologies outlined in this work---provides a firm basis for spectroscopic studies of liquid water (such as X-ray absorption and photo-electron spectroscopy),~\cite{kong_prb_2012,swartz_prl_2013} as well as the study of solvation structure in aqueous ionic solutions, which play a key role in biology and energy research.

Although the simulations performed in this work provide us with a more accurate description of the microscopic structure of liquid water, an explicit treatment of the NQE (\textit{i.e.}, \textit{via} the Feynman discretized path integral (PI) approach) is still lacking in our current approach and will be required to quantitatively describe certain structural properties that directly involve the lighter hydrogen atoms, which can significantly deviate from classical behavior even at room temperature.
The explicit inclusion of NQE \textit{via} PI-AIMD simulations coupled with an underlying XC functional that accounts for both Exx and vdW interactions provides an accurate and balanced theoretical treatment of both the nuclear and electronic degrees of freedom in liquid water; as such, this is the focus of intense ongoing research in our group and will be addressed in future work.

Additional future work should also be devoted to addressing remaining deficiencies in the underlying XC functional by further reducing the deleterious effects of electron self-interaction. 
In addition, the theoretical description of the vdW/dispersion interactions could also be improved \textit{via} the inclusion of beyond-pairwise interactions, \textit{e.g.}, as provided by the many-body dispersion (MBD) framework.~\cite{tkatchenko_prl_2012,distasio_pnas_2012,tkatchenko_jcp_2013,ambrosetti_jcp_2014,distasio_jpcm_2014}
%
%
%
Moving beyond the realm of DFT, the theoretical description of the electronic degrees of freedom could also be addressed using more accurate quantum chemistry approaches and/or quantum Monte Carlo; however, the high computational cost associated with these methodologies has restricted their use to date in large-scale AIMD simulations of condensed-phase systems.
%

\begin{acknowledgments}
R.D., B.S., and R.C. acknowledge support from the Department of Energy (DOE) under Grant Nos. DE-SC0005180 and DE-SC000826.
R.D., Z.L., and R.C. also acknowledge support from the National Science Foundation (NSF) under Grant No. CHE-0956500 during the early stages of this work. 
X.W. acknowledges support from the American Chemical Society Petroleum Research Fund (ACS PRF) under Grant No. 53482-DNI6 and the DOE under Grant No. DE-SC0008726.
This research used resources of the National Energy Research Scientific Computing Center, which is supported by the Office of Science of the U.S. Department of Energy under Contract No. DE-AC02-05CH11231.
This research used resources of the Argonne Leadership Computing Facility at Argonne National Laboratory, which is supported by the Office of Science of the U.S. Department of Energy under contract DE-AC02-06CH11357.
Additional computational resources were provided by the Terascale Infrastructure for Groundbreaking Research in Science and Engineering (TIGRESS) High Performance Computing Center and Visualization Laboratory at Princeton University.
\end{acknowledgments}

\bibliography{water}

\begin{thebibliography}{115}
\bibitem [{coo()}]{coordination_number_footnote}%
  \BibitemOpen
  \href@noop {} {}\bibinfo {note} {{The quantity $n_{\rm OO}$ is very sensitive
  to $r^{\rm min}_{1}$, the position of the first minimum in the $g_{\rm
  OO}(r)$, since $n_{\rm OO}$ is computed as $n_{\rm OO}=4\pi\rho
  \int_{0}^{r^{\rm min}_{1}} dr\,r^{2}g_{\rm OO}(r)$, with $\rho$ as the
  corresponding atomic number density.}}\BibitemShut {Stop}%
\bibitem [{hbo()}]{hbond_footnote}%
  \BibitemOpen
  \href@noop {} {}\bibinfo {note} {{To obtain the number of distinct hydrogen
  bonds per water molecule, the numbers provided in Table~\ref{tab1} should be
  divided by two.}}\BibitemShut {Stop}%
\bibitem [{mlw()}]{mlwf_footnote}%
  \BibitemOpen
  \href@noop {} {}\bibinfo {note} {{A hydrogen bond is assigned to an MLWF if
  the distance between the center of the MLWF ($W_c$) and the donating hydrogen
  atom is less than 2.7~\AA\ and the angle $\angle {\rm W_c\!\!-\!\!O \cdots
  H}$ is less than 30$^\circ$. For a given hydrogen bond, if both of the MLWFs
  satisfy these two geometric criteria (either partially or fully), then the
  two MLWFs share that particular hydrogen bond with a distance-weighted
  fraction.}}\BibitemShut {Stop}%
\end{thebibliography}

\end{document}